\documentclass[preprint,12pt]{aastex}
\usepackage{emulateapj5}

\newcommand{\zform}{z_{\rm form}}
\newcommand{\micr}{$\mu$m}
\newcommand{\sqdeg}{deg$^2$}
\newcommand{\bj}{$b_{\rm J}$}
\newcommand{\cosmos}{$(\Omega_{m_0},\Omega_{\Lambda_0})$}
\newcommand{\cosmoparas}{$(h,\Omega_{m_0},\Omega_{\Lambda_0})$}
\newcommand{\dd}{{\rm d}}
\newcommand{\Msun}{\hbox{M$_{\odot}$}}
\newcommand{\Lsun}{\hbox{L$_{\odot}$}}
\newcommand{\Zsun}{\hbox{Z$_{\odot}$}}
\newcommand{\rr}{$^{0.1}r$}
\newcommand{\sfrunits}{$h\,{\rm M}_{\odot}\,{\rm yr}^{-1}\,{\rm Mpc}^{-3}$}
\newcommand{\bolounits}{$h\,{\rm W}\,{\rm Mpc}^{-3}$}
\newcommand{\rhosfr}{$\rho_{\rm SFR}$}
\newcommand{\omstars}{$\Omega_{\rm stars}$}

\begin{document}

\title{Constraints on a Universal IMF 
  from UV to Near-IR Galaxy Luminosity Densities}
\shorttitle{Constraints on an IMF from Luminosity Densities}
\author{Ivan~K.~Baldry and Karl~Glazebrook}
\shortauthors{I.~K.~Baldry \& K.~Glazebrook}
\affil{Department of Physics \& Astronomy, Johns Hopkins University,
  Baltimore, MD~21218-2686, USA}

\begin{abstract}
  We obtain constraints on the slope of a universal stellar initial
  mass function (IMF) over a range of cosmic star-formation histories
  (SFH) using $z\approx0.1$ luminosity densities in the range from
  0.2\,\micr\ to 2.2\,\micr.  The age-IMF degeneracy of integrated
  spectra of stellar populations can be broken for the Universe as a
  whole by using direct measurements of (relative) cosmic SFH from
  high-redshift observations. These have only marginal dependence on
  uncertainties in the IMF, whereas, fitting to local luminosity
  densities depends strongly on both cosmic SFH and the IMF.  We fit
  to these measurements using population synthesis and find the
  best-fit IMF power-law slope to be $\Gamma=1.15\pm0.2$ (assuming
  $\dd N / \dd \log m\,\propto\, m^{-\Gamma}$ for 0.5--120\,\Msun\ and
  $m^{-0.5}$ for 0.1--0.5\,\Msun). This $M>0.5$\,\Msun\ slope is in
  good agreement with the Salpeter IMF slope ($\Gamma=1.35$). A strong
  upper limit of $\Gamma<1.7$ is obtained which effectively rules out
  the Scalo IMF due to its too low fraction of high-mass stars.  This
  upper limit is at the 99.7\% confidence level if we assume a
  closed-box chemical evolution scenario and 95\% if we assume
  constant solar metallicity.  Fitting to the H$\alpha$ line
  luminosity density, we obtain a best-fit IMF slope in good agreement
  with that derived from broadband measurements.

  Marginalizing over cosmic SFH and IMF slope, we obtain (95\% conf.\ 
  ranges): \omstars\ = 1.1--2.0 $\times10^{-3}$ $h^{-1}$ for the
  stellar mass density; \rhosfr\ = 0.7--4.1 $\times10^{-2}$ \sfrunits\ 
  for the star-formation rate density, and; $\rho_{L}$ = 1.2--1.7
  $\times10^{35}$ \bolounits\ for the bolometric, attenuated, stellar,
  luminosity density (0.09--5\,\micr).  Comparing this total stellar
  emission with an estimate of the total dust emission implies a
  relatively modest average attenuation in the UV ($\la1$ magnitude at
  0.2\,\micr).
\end{abstract}


\section{Introduction}
\label{sec:intro}

The luminosity density of the Universe in the range from about
0.2\,\micr\ to 2.2\,\micr, mid-UV to near-IR, is dominated by stellar
emission. This wavelength range spans the peak in the spectra
($f_\nu$) of stars with effective temperatures from about 20000\,K
down to 2000\,K.  Therefore, measurements of the luminosity density in
various broadbands across this range provide a powerful constraint on
cosmic star-formation history (SFH) and/or a universal stellar initial
mass function (IMF).

The stellar IMF describes the relative probability of stars of
different masses forming \citep[see][for recent reviews and
analyses]{GH98}. Its importance crosses many fields of astronomy from,
for example, star formation (testing theoretical models) to cosmic
chemical evolution (heavy metal production from high mass stars).  It
is widely used in the study of the SFH of galaxies from their
integrated spectra. Generally, an IMF is assumed and used as an input
to evolutionary, stellar population, synthesis models, and these
models are fitted to integrated spectra.  

The first calculation of an IMF was made by \cite{Salpeter55} based on
the observed luminosity function of solar-neighborhood stars,
converting to mass, correcting for main-sequence lifetimes and
assuming that the star-formation rate (SFR) has been constant for the
last 5\,Gyr.  Despite the uncertainties in mass-to-light ratios,
stellar lifetimes and the SFR, this result (a power-law slope of
$-1.35$ measured from about 0.3 to 15\,\Msun) is still commonly used
today.\footnote{The Salpeter IMF power-law slope is $-1.35$ wrt.\ 
  logarithmic mass bins and $-2.35$ wrt.\ linear mass bins.}

\begin{figure*}[ht] 
\epsscale{1.0}\plotone{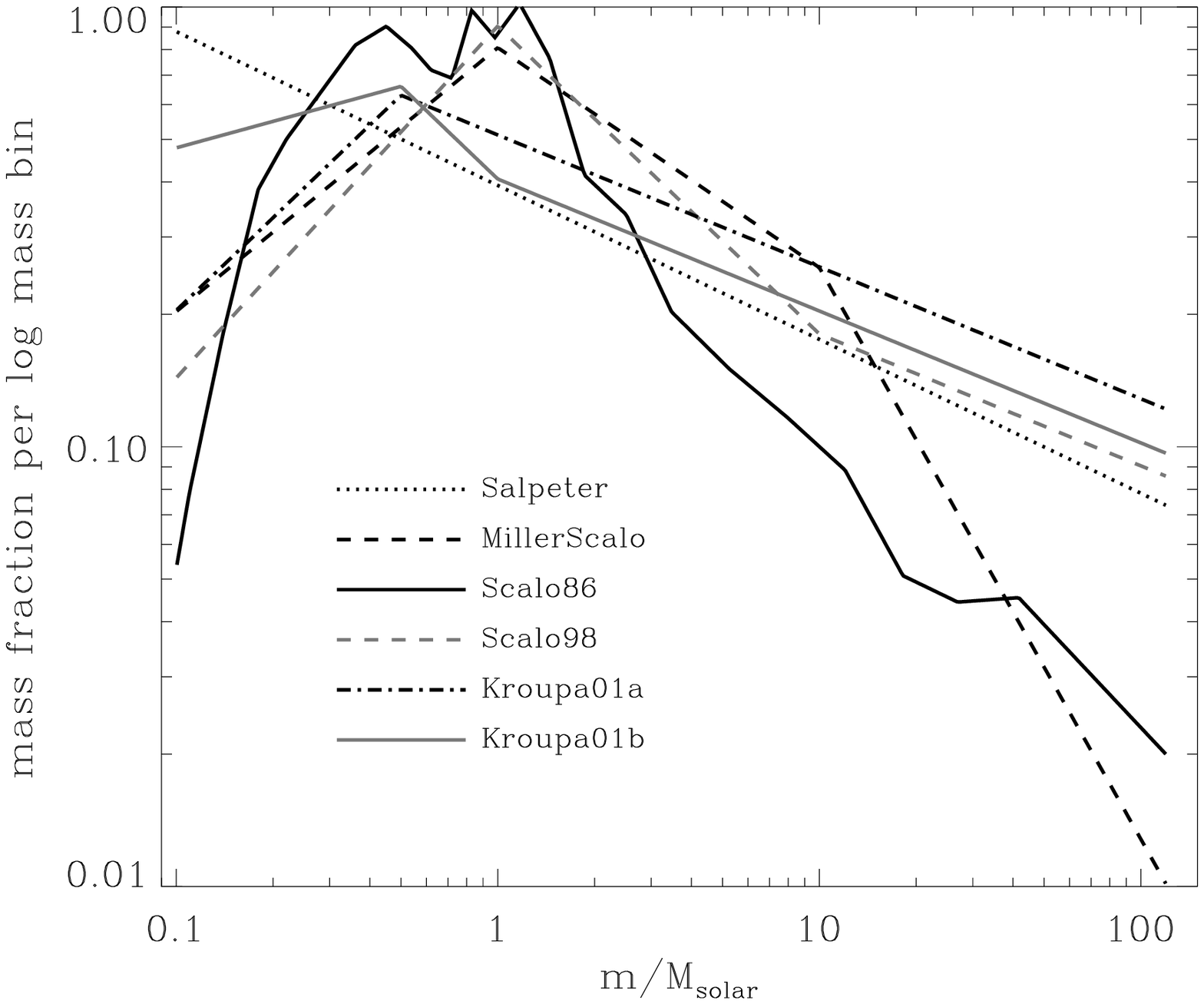}
\epsscale{1.0}\plotone{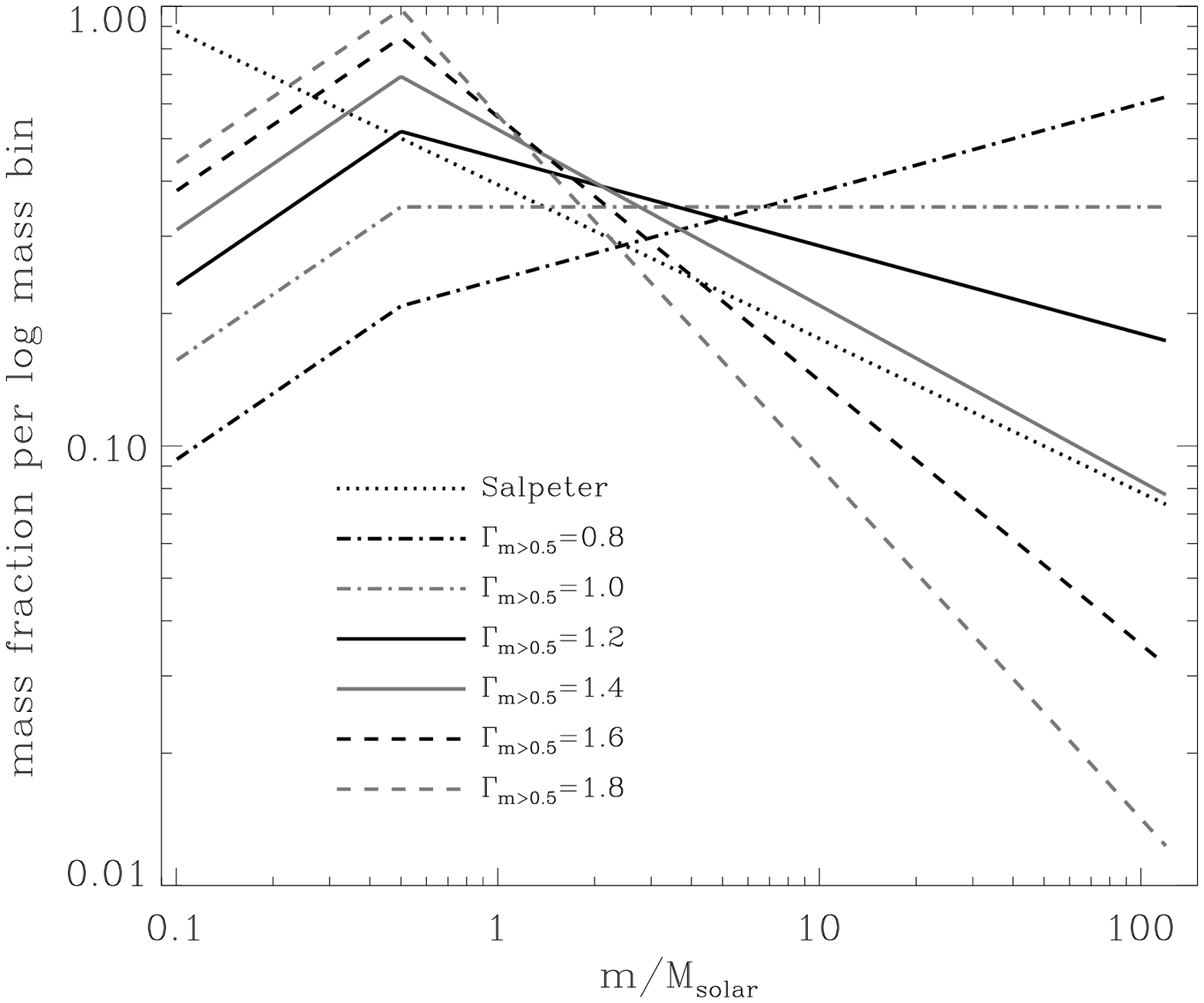}
\caption{Stellar initial mass functions:{} 
  mass fraction (per logarithmic mass bin) versus mass for the
  \cite{Salpeter55} IMF, the \cite{MS79} IMF, the
  \cite{Scalo86,Scalo98} IMFs, the \citeauthor{Kroupa01}
  (\citeyear[eqns.\ 2 and 6]{Kroupa01}) IMFs (left panel) and the IMF
  parameterization of Equation~\ref{eqn:imf} (right panel).  All are
  assumed to be valid over the range 0.1--120\,\Msun.  The integral of
  each curve is set to unity.  Note that the slopes of the lines are
  equivalent to $1-\Gamma$ [$\,=\dd\log(m\,n_{\log
    m})\,/\,\dd\log(m)$, see Eqn.~\ref{eqn:imf} for definitions]. The
  $\Gamma$ values for $M>1$\Msun\ for the published IMFs are 1.35
  (S55), 1.5/2.3 (MS79), $\sim$2.05/1.5 (S86), 1.7/1.3 (S98) and 1.3
  (K01) (see also Table~\ref{tab:publ-imfs}).}
\label{fig:imfs}
\end{figure*}

Measurements of the solar neighborhood IMF were reviewed by
\cite{Scalo86} producing an IMF with a mass fraction peak around
0.5--1\,\Msun\ (Fig.~\ref{fig:imfs}). When applied to galaxy
populations, this IMF is unable to reproduce H$\alpha$ luminosities
\citep*{KTC94}, and applied to cosmic evolution, is unable to match
observed mean galaxy colors \citep*{MPD98} due to a too low fraction
of high-mass stars ($M>10$\,\Msun). 

In a more recent review by \cite{Scalo98}, he concluded that the field
star IMF was of questionable use for a number of reasons (e.g.\ the
derived IMF in the range 0.9--1.4\,\Msun\ depends strongly on the
assumed solar-neighborhood SFH).  Instead, he summarized the results
from studying star clusters in a triple-index power-law IMF as an
estimate of an average IMF (Fig.~\ref{fig:imfs}).  He also noted that
``if the existing empirical estimates of the IMF are taken at face
value, they present strong evidence for variations, and these
variations do not seem to depend systematically on physical variables
such as metallicity or stellar density''.  The observed IMF variations
from stellar counts could be largely due to statistical fluctuations
and/or observational biases such as mass segregation within a star
cluster \citep{Elmegreen99}.  These points, if correct, mean that the
concept of a universal IMF is highly useful in the studies of
integrated spectra of galaxies (since they are mostly the result of
many star-formation regions and episodes) but the IMF is still
uncertain at the level of 0.5 in the power-law slopes.  By `universal
IMF', we mean an IMF that represents the average in a significant
majority of galaxies and over a significant majority of cosmic time.

Measurements of the cosmic luminosity densities represent the
$\lambda/\Delta\lambda\sim6$ components of the ``Cosmic Spectrum''
which is the luminosity-weighted, average galaxy spectrum
\citep{baldry02}.  Thus, if there is a universal IMF, it should
certainly apply to this spectrum and constraints on the IMF can be
obtained by fitting population synthesis models.  These constraints
are degenerate with the assumed cosmic SFH.  However, cosmic SFH is
generally more accurately quantified than the SFH of any individual
galaxy other than in the Local Group.  For example, star-forming
galaxies have been observed as far back as $z\sim5$ \citep{dey98}, and
there is evidence for reionization as early as $z\sim20$
\citep{kogut03}, which for a Universe age of $13.7\pm0.2$\,Gyr
\citep{spergel03} gives a time of 12.3--13.7\,Gyr since the onset of a
significant rate of star formation in the Universe.  The results for
individual galaxies are often limited by age-metallicity degeneracy
\citep{Worthey94} and recent bursts of star formation which can
disguise their underlying age \citep{BG97}.  Given our knowledge of
cosmic SFH, we can then place constraints on a universal IMF by
limiting the SFH we consider.

The primary knowledge of cosmic SFH comes from measuring the comoving
density of SFR indicators at various redshifts \citep{madau96}. These
SFR indicators include UV luminosities and emission-line luminosities
that are dominated by light from short-lived high-mass stars. The
conversion to SFR depends on the IMF.  However, the {\em relative}
cosmic SFH is well defined if the same indicator is used at each
redshift regardless of the assumed IMF.  Even if the indicator varies
(e.g.\ 0.2 to 0.3\,\micr\ rest-frame UV), the derived cosmic SFH is
less sensitive to the slope of the IMF than the local luminosity
densities (0.2 to 2.2\,\micr).  It is this principal that enables
constraints on a universal IMF.

An analysis of this type was applied by \cite{MPD98} using various
luminosity densities (0.15 to 2.2\,\micr) spread over a range of
redshifts (0--4). They fitted cosmic SFHs for three different IMFs.
Here, we use a more quantitative approach to constraining a universal
IMF slope and use more accurate, recent, local ($z\approx0.1$)
luminosity density measurements. Note that we assume that there is a
universal IMF and do not constrain any variation in the IMF between
different galaxies \citep[see e.g.][for evidence for an invariant
IMF]{Wyse97,Kroupa02}.  Even if there is some variation, the results
presented here could be regarded as constraints on a
luminosity-weighted, average IMF.

To summarize, if the cosmic SFH is assumed to be known, based on
measurements of various SFR indicators with redshift, then local
luminosity density measurements provide a constraint on a universal
stellar IMF.  The other principal factors to consider are chemical
evolution (metallicity) and dust attenuation.

\begin{figure*} 
\epsscale{1.0}\plotone{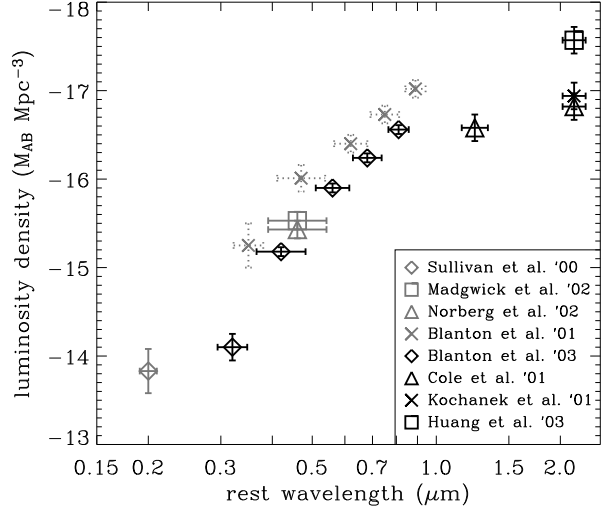}
\caption{Local luminosity densities:{} 
  absolute AB magnitudes per Mpc$^3$ [\cosmoparas\ = (1.0,0.3,0.7)]
  versus wavelength.  The vertical bars represent uncertainties while
  the horizontal bars represent FWHM of the bandpasses. See
  Table~\ref{tab:lum-dens} and Section~\ref{sec:lum-dens} for details.
  We use the \cite{sullivan00}, the \cite{blanton03} and the
  $J$-$K$-band results for our fitting.}
\label{fig:lum-dens}
\end{figure*}

The plan of the paper is as follows. In Section~\ref{sec:lum-dens}, we
describe details of recent luminosity density measurements. The
measurements are illustrated in Figure~\ref{fig:lum-dens} and
summarized in Table~\ref{tab:lum-dens}. In Section~\ref{sec:modeling},
we describe our modeling and fitting procedure. A summary of the
parameters used in the modeling is given in
Section~\ref{sec:summary-paras}. In Sections~\ref{sec:results}
and~\ref{sec:conclusions}, we present our results and conclusions.

\begin{table*} 
\caption{Local luminosity densities from various surveys{} ($0<\bar{z}<0.2$)}
\label{tab:lum-dens}
\begin{center}
\begin{tabular}{llccl} \hline
band & reference & $\lambda_{\rm eff}$ &$j+2.5\log h$ & notes \\
     &           &  (\micr)            &\llap{$^a$}(AB magnitudes)& \\ \hline
\llap{$^b$}FOCA 0.2\,\micr & \cite{sullivan00} & 0.20 & $-13.83 \pm 0.25$ & $z\approx0.15$, uncorrected for dust \\
\llap{$^c$}APM \bj      & \cite{madgwick02} & 0.46 & $-15.53 \pm 0.10$ & $z\approx0.10$ \\
''           & \cite{norberg02}  & 0.46 & $-15.43 \pm 0.10$ & evolution corrections to $z=0.0$ \\
\llap{$^d$}SDSS $u$     & \cite{blanton01}  & 0.35 & $-15.25 \pm 0.25$ & $z\approx0.05$ \\
SDSS $g$     & ''                & 0.47 & $-16.01 \pm 0.15$ & $z\approx0.10$ \\
SDSS $r$     & ''                & 0.62 & $-16.40 \pm 0.10$ & '' \\
SDSS $i$     & ''                & 0.75 & $-16.73 \pm 0.10$ & '' \\
SDSS $z$     & ''                & 0.89 & $-17.02 \pm 0.10$ & '' \\
SDSS $^{0.1}u$ & \cite{blanton03}& 0.32 & $-14.10 \pm 0.15$ & evolution corrections to $z=0.1$ \\
SDSS $^{0.1}g$ & ''              & 0.42 & $-15.18 \pm 0.05$ & '' \\
SDSS $^{0.1}r$ & ''              & 0.56 & $-15.90 \pm 0.05$ & '' \\
SDSS $^{0.1}i$ & ''              & 0.68 & $-16.24 \pm 0.05$ & '' \\
SDSS $^{0.1}z$ & ''              & 0.81 & $-16.56 \pm 0.05$ & '' \\
\llap{$^e$}2MASS $J$    & \cite{cole01}     & 1.24 & $-16.58 \pm 0.15$ & evolution corrections to $z=0.0$ \\
2MASS $K_s$  & ''                & 2.16 & $-16.82 \pm 0.15$ & evolution corrections to $z=0.0$ \\
''           & \cite{kochanek01} & 2.16 & $-16.94 \pm 0.15$ & $z\approx0.03$ \\
Hawaii $K_s$ & \cite{huang03}    & 2.16 & $-17.57 \pm 0.15$ & $z\approx0.15$ \\ \hline
\end{tabular}
\end{center}
$^a$luminosity density conversion: $j = -2.5 \log( L_{\nu} / {\rm \,W\,Hz^{-1}\,Mpc^{-3}} ) + 34.10$\newline
$^b$FOCA magnitudes: assumed $m_{\rm AB} = m_{2000} + 2.25$ 
  \citep{milliard92}\newline
$^c$APM magnitudes: assumed $m_{\rm AB} = b_{\rm J} - 0.08$
  (see Section~\ref{sec:ld-optical})\newline
$^d$SDSS magnitudes: assumed $m_{\rm AB} = 
  (u,g,r,i,z)\,+\,(-0.04,0.035,0.015,0.015,0.00)$ \citep{blanton03}\newline
$^e$2MASS, Hawaii magnitudes: assumed $m_{\rm AB} = J+0.9$, 
  $m_{\rm AB} = K+1.8$ \citep{cutri01}
\end{table*}

\section{Luminosity Density Measurements}
\label{sec:lum-dens}

In this section, we summarize the details of various luminosity
density measurements and their conversion to AB magnitudes
\citep{OG83} per comoving Mpc$^3$.  The zero point of the AB {\em
  absolute} magnitude scale is $4.345 \times 10^{13}\: {\rm W}\: {\rm
  Hz}^{-1}$ (3631\,Jy for {\em apparent} magnitudes).  In general, an
estimate of the total luminosity density from the galaxy population is
obtained by an analytical integration of the \cite{Schechter76}
function parameters, given by
\begin{equation}
j = M_{*} - 2.5 \log 
  \left[ \phi_{*} \, \Gamma_{\rm f} (\alpha + 2) \right] + {\cal C}
\label{eqn:lum-dens-schechter}
\end{equation}
where $\Gamma_{\rm f}$ is the gamma function and ${\cal C}$ represents
corrections from the magnitude system defining the luminosity function
to total AB magnitudes. The final luminosity densities are quoted for
a cosmology where \cosmoparas\ = (1.0,0.3,0.7) and $h = H_0 /
100\,{\rm km\,s^{-1}\,Mpc^{-1}}$.  The surveys select redshifts to a
limiting magnitude in the same wavelength as that of the luminosity
density measurement. 

\subsection{Ultraviolet: 0.2\,\micr}
\label{sec:ld-uv}

An analysis of a mid-UV selected redshift survey is presented by
\cite{sullivan00}. The imaging for this survey covers about 6\,\sqdeg\
to a depth of about 21 AB magnitudes using the FOCA balloon-borne
telescope \citep{milliard92}.  The filter response approximates a
Gaussian centered at 2015\,\AA\ with a full-width half maximum (FWHM) of
188\,\AA.  The survey area covers four fields chosen with very low
milky-way (MW) extinction [$E(B-V)\le0.015$].  Follow up spectroscopy
was obtained using multi-object, optical spectrographs on the WIYN and
WHT telescopes \citep{treyer98}.

\citeauthor{sullivan00}\ measured the local luminosity density using a
sample of about 200 galaxies ($0<z<0.4$) to estimate the luminosity
function. The \citeauthor{Schechter76} parameters without
dust-correction were $(M_{*},\phi_{*},\alpha) =
(-20.59,0.00955,-1.51)$. Integrating this function gives $-16.18$, and
converting to AB magnitudes gives $-13.93$ using $m_{\rm AB} =
m_{2000} + 2.25$ \citep{milliard92}.  We also correct the measurement
to our default world model of \cosmos\ = (0.3,0.7) from (1.0,0.0)
assuming a redshift of 0.15.  This gives a correction of about $+0.1$
magnitudes.  The formal uncertainty from the fitting is 0.13
magnitudes. However, there are additional uncertainties: absolute
calibration ($<0.25$ mags), MW extinction ($<0.15$ mags), conversion
to total magnitudes ($<0.1$ mags), large-scale structure,
incompleteness corrections, etc.  We will assume an additional
1$\sigma$ uncertainty of 0.2 to be added in quadrature, so that
$j=-13.83\pm0.25$.

The limit of this survey is $m_{2000}=18.5$ which corresponds to
$M_{*}$ galaxies at $z\approx0.2$. With a steep faint-end slope of
$-1.5$, the luminosity-weighted mean redshift is around 0.15
corresponding to the redshift position plotted by
\citeauthor{sullivan00} This is higher than our fiducial redshift of
0.10 (see below), giving a higher luminosity for any declining cosmic
SFR at $z<0.5$.  However, this may be counteracted by the lack of
MW-extinction and total-magnitude corrections.  For simplicity and
since we do not want to assume a cosmic SFH and IMF, we will use the
luminosity density as it was measured. Note that a couple of galaxies
with obvious active galactic nuclei (AGN) characteristics were removed
from their sample.  We will assume the measured luminosity density
does not have a strong, non-stellar, AGN component. Neither
\cite{sullivan00} or \cite{contini02} found strong evidence for
significant AGN contamination based on emission-line flux ratios.

\subsection{Optical: 0.3\,\micr\ to 0.9\,\micr}
\label{sec:ld-optical}

The Sloan Digital Sky Survey \citep[SDSS;][]{york00,stoughton02} has
imaged over 2000\,\sqdeg\ in five bandpasses ($ugriz$) with effective
wavelengths from 0.35 to 0.9\,\micr\ to a depth of about 21--22 AB
magnitudes. Followup spectroscopy is also included as part of the SDSS
for various targeting schemes, of which, the main galaxy sample
\citep[MGS;][]{strauss02} is appropriate for determining cosmic
luminosity densities.  The MGS is a magnitude-limited galaxy sample
($r<17.77$) with a median redshift of 0.10. To form effectively
complete samples in the other four bands, galaxies were selected to
$(u,g,i,z)\:\la\:(18.4,17.65,16.9,16.5)$.

The first luminosity densities from the MGS were published by
\cite{blanton01}. However, \cite{Wright01} found that these results
overpredicted the near-IR luminosity density by a factor of 2.3
\citep[compared to][]{cole01,kochanek01}.  Since then, better analysis
techniques, calibration and more data have allowed the optical
luminosity density measurements to be significantly improved.  We use
the results of \cite{blanton03} as the basis for our fitting here. One
of their approaches was to $k$-correct and to evolve-correct to a
fiducial redshift of 0.10.  This reduces systematic uncertainties
associated with these types of corrections because the median redshift
of the SDSS is at this mark and the luminosity-weighted mean redshift
is close to it. Thus, their results are most accurate for the shifted
bandpasses, designated $^{0.1}u$, $^{0.1}g$, $^{0.1}r$, $^{0.1}i$ and
$^{0.1}z$ (rest-frame bandpasses for galaxies at $z=0.1$). We use the
\rr\ band as the fiducial band and measure all colors with respect to
it when comparing synthetic magnitudes with luminosity densities.  We
set a 1$\sigma$ uncertainty of 0.05 for the $g,r,i,z$ band
measurements.  This allows for some mis-calibration since the formal
uncertainties of \citeauthor{blanton03}\ are 0.03/0.02 for these
bands. Using such small errors is appropriate since it is the relative
$^{0.1}g,^{0.1}i,^{0.1}z$ to \rr\ measurements that constrain the
normalized SFH or IMF. In other words, the errors only need to
represent the uncertainties in the colors and the absolute
measurements of the luminosity densities need not be accurate to this
level. An estimated conversion of SDSS to AB magnitudes is given and
used by \citeauthor{blanton03}\ (see also Table~\ref{tab:lum-dens}),
and the magnitudes used by them are assumed to be close enough to
total that a correction is not applied.

The new results of \cite{blanton03} are in good agreement with \bj\ 
luminosity densities determined by \cite{madgwick02} and
\cite{norberg02}.  These analyses were based on the Automated Plate
Measuring \citep[APM;][]{maddox90} galaxy catalog with redshifts from
the 2dF Galaxy Redshift Survey \citep[2dFGRS;][]{colless01}.  The
conversion to AB magnitudes of $-0.08$ was based on an integration of
the \bj\ curve (P.~C.\ Hewett \& S.~J.\ Warren 1998 private
communication) through a spectrum of Vega \citep*[computed by R.~L.\ 
Kurucz]{LCB97}. The \bj\ magnitudes were calibrated to total by
comparison with deeper, CCD photometry.

\subsection{Infrared: 1.0\,\micr\ to 2.5\,\micr}
\label{sec:ld-ir}

The Two Micron All Sky Survey \citep[2MASS;][]{skrutskie97} has imaged
the whole sky in the $J$, $H$ and $K_s$ bands to a depth of 15--16 AB
magnitudes (14.7, 13.9, 13.1 Vega mags, respectively).  \cite{cole01}
matched the second incremental data release to redshifts obtained by
the 2dFGRS.  With this data set, they determined the local $J$ and $K$
band luminosity functions.  With $k$ and evolution corrections to
$z=0$, the \citeauthor{Schechter76} parameters were
$(-22.36,0.0104,-0.93)$ and $(-23.44,0.0108,-0.96)$ for the $J$ and
$K$ bands, respectively. Integrating these functions gives $-17.36$
and $-18.50$. The conversions to AB magnitudes are taken as $m_{\rm
  AB} = J+0.9$ and $K+1.8$ \citep{cutri01} and the conversion to total
magnitudes is estimated to be between $-0.08$ and $-0.15$ (we take
$-0.12$).  The uncertainties come from Poisson noise, the absolute
calibration, the conversion to total magnitudes and large-scale
structure.  These are not all well defined but we will be conservative
and use 0.15 for the 1$\sigma$ uncertainty so that $j=-16.58\pm0.15$
and $j=-16.82\pm0.15$ for the $J$ and $K$ bands, respectively.

The $K$-band luminosity function was also determined by
\cite{kochanek01} using 2MASS imaging. Here, they determined the
luminosity density using a shallower sample but with greater sky
coverage for the redshifts.  Their best estimate was obtained by
summing separate luminosity functions for late and early-type
galaxies: $(-22.98,0.0101,-0.87)$ and $(-23.53,0.0045,-0.92)$. The
total luminosity density is then $-16.94$ in AB magnitudes after
appropriate corrections (to AB, as above; to total, $-0.21$). This is
in good agreement with the \citeauthor{cole01}\ result.

A deeper $K$-band survey covering about 8\,\sqdeg\ was recently
analysed by \cite{huang03}.  Imaging was taken with the University of
Hawaii telescopes at Mauna Kea Observatory \citep{huang97} with
redshifts obtained using the 2dF facility on the AAT. The best-fit
\citeauthor{Schechter76} parameters were $(-23.70,0.0130,-1.37)$,
giving a luminosity density of $j=-17.57$ after correcting to AB
magnitudes.  Even with a conservative error of 0.15, this result is
$>3\sigma$ discrepant from the result of \cite{cole01}. This
discrepancy of 0.75 magnitudes is too large to be explained by cosmic
evolution since the $K$-band luminosity density is dominated by the
older stellar populations. Analysis by \cite{huang03} suggests it
could be due to a local underdensity. We note that near-IR luminosity
densities have larger uncertainties due to large-scale structure than
bluer measurements. In underdense regions, the SFR per galaxy is
higher \citep[as noted by color-density and morphology-density
relationships, e.g.][]{blanton03broadband} which counteracts the
effect on the UV luminosity density.

Given the above discrepancy in the near-IR luminosity densities,
firstly, we will consider the \citeauthor{cole01}\ results separately
from the \citeauthor{huang03}\ results in our fitting, and secondly,
we will use an average $K$-band luminosity.

\section{Modeling}
\label{sec:modeling}

\begin{figure*}[ht] 
\epsscale{1.0}\plotone{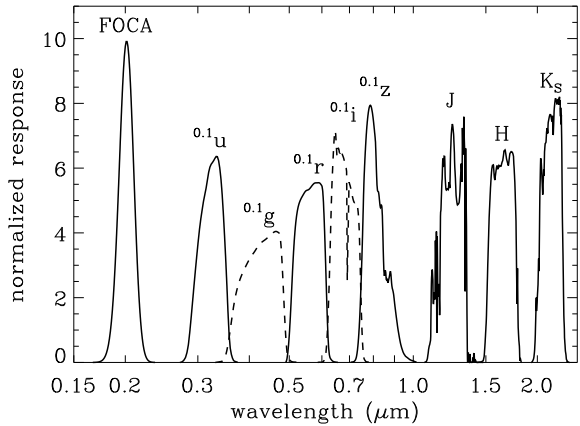}
\caption{FOCA, SDSS and 2MASS filters.
  The SDSS filters are shown `shifted' to $z=0.1$ since we are fitting
  to the results of \cite{blanton03}.  Each curve is normalized so
  that the integral of the normalized transmission $T\,\dd\ln\lambda$
  is equal to unity.  Thus the height of each curve is related to its
  resolving power ($\lambda/\Delta\lambda$).}
\label{fig:filters}
\end{figure*}

We fit the data with synthetic magnitudes calculated using the
PEGASE.2 evolutionary synthesis code \citep{FR97,FR99} and integrating
the spectra through the filter response curves
(Fig.~\ref{fig:filters}).  The responses for the SDSS are taken from
\cite{stoughton02}, the FOCA filter curve is assumed to be Gaussian
\citep{milliard92} and the near-IR filter curves are taken from
\cite{cutri01}. We run the PEGASE models with nebular emission but `no
extinction' except we modified the code so that the absorption of
Lyman continuum photons by dust was still included \citep[following
the prescriptions of][]{Spitzer78}.  This was done because our dust
attenuation parameterization (Sec.~\ref{sec:model-dust}) does not
apply to this `pre-extinction' of nebular continuum and line emission.
The prescription for the models is described below including SFH, IMF
and chemical evolution.  We include further effects of dust
attenuation on the output spectra.

\subsection{Cosmic SFH}
\label{sec:model-sfh}

For our modeling of cosmic star-formation history we use a `double
power-law' parameterization \citep{baldry02}:
\begin{equation} {\rm SFR} \: \propto
\begin{array}{ll}
(1+z)^\beta   &   \hbox{  for } 0<z<1 \\
(1+z)^\alpha  &   \hbox{  for } 1<z<5 \\
\end{array}
\label{eqn:cosmic-sfh}
\end{equation}
The SFRs from the two power-laws are the same at $z=1$ and star
formation is started at $z=5$. Clearly, there are many other possible
cosmic SFHs other than defined by these two parameters. However, the
resulting synthetic spectra are highly degenerate with SFH.  We chose
this two-parameter model because the $\beta$ parameter provides a good
match to direct measures of SFR at $z\la1$ and the $\alpha$ parameter
allows us to add high-redshift star formation in a well-defined way.
We assume a cosmology corresponding to \cosmoparas = (0.7,0.3,0.7).
Most importantly this determines the timescale for cosmic SFH
measurements.

\begin{figure*}[ht] 
\epsscale{1.0}\plotone{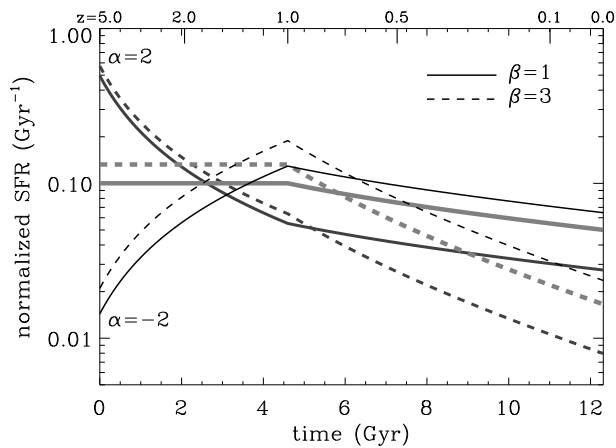}
\caption{Examples of the parameterized cosmic SFH.
  The SFR is proportional to $(1+z)^\beta$ for $z<1$ and
  $(1+z)^\alpha$ for $1<z<5$. Each curve is normalized so that the
  total star formation between $z=5$ and $z=0$ is unity. The timescale
  is for \cosmoparas = (0.7,0.3,0.7). Six SFHs are shown with two
  values of $\beta$ (1,3) for each of three values of $\alpha$
  ($-$2,0,2).}
\label{fig:sfhs}
\end{figure*}

Examples of this parameterized cosmic SFH are shown in
Figure~\ref{fig:sfhs} with a timescale for the $h=0.7$ cosmology.
Note that direct measures of the {\em relative} cosmic SFH with
redshift do not depend on $H_0$. The Hubble constant changes the
scaling factor at all redshifts by the same amount. However, our
modeling of the cosmic SFH, to fit to local luminosity densities (fossil
cosmology), does depend on $H_0$ because of the timescale dependence.
The synthetic spectra are calculated at 11\,Gyr corresponding to the
fiducial redshift, $z=0.1$, with the chosen cosmology and $\zform=5$.

There is overwhelming evidence for a rise in SFR to $z=1$ from a
variety of indicators
\citep[e.g.][]{haarsma00,hammer97,lilly96,rowan-robinson97}.  Recent
estimates from compilations of measures of luminosity density with
redshift out to $z=1$ give, for example, $\beta=2.7\pm0.7$
\citep{Hogg02}.  If we take the formal 2$\sigma$ range from
\citeauthor{Hogg02}, we obtain a range of 1.3--4.1.  However, some of
the measurements compiled by him could be biased toward a steeper
evolution (e.g.\ \citeauthor{lilly96}) because of selection effects.
\cite*{CSB99} found a shallower evolution of $\beta=1.5\pm0.5$ based
on rest-frame UV selection at all redshifts.  To encompass the
``evidence for a gradual decline'' (\citeauthor{CSB99}) to a steep
decline (\citeauthor{lilly96}), we will consider $\beta$ to be in the
range 0.5--4.0.

High redshift ($z>1$) star formation is parameterized by $\alpha$ and
$\zform=5$. We do not try to exactly match high redshift measurements.
The uncertainties are still large due to, for example, dust and
surface brightness corrections. By taking $\alpha$ from $-2$ to $2$,
we can approximate the effect of high-redshift star formation from a
rapid decline at high redshift \citep{madau96} to a flattening
\citep{steidel99} to a significant rise \citep{lanzetta02}.  Note that
the effect of any significant star formation in the $\sim$1\,Gyr
between $z=20$ and $z=5$ can be approximated by an increase in
$\alpha$.

\subsection{Universal IMF}
\label{sec:model-imf}

Significant degeneracies exist with modeling the spectra of galaxies,
e.g.\ age-metallicity.  Given these uncertainties, modelers have
generally assumed an IMF, typically the \cite{Salpeter55} IMF for
modeling galaxy spectra. Here, we will assume a universal IMF
(constant with time and environment) but we will consider different
IMF power-law slopes. A double power-law IMF is chosen based on the
rationale that there is a clear change in slope around 0.5\,\Msun\ but
no definitive change at higher masses \citep{Kroupa01}. Our
parameterization is\footnote{An equivalent formalism{} for our
  upper-mass slope is $\,\dd N / \dd \log m\,\propto\, m^{-\Gamma}\,$
  where $N$ is the cumulative number of stars up to mass $m$. This
  follows the nomenclature outlined by \cite{Kennicutt98imf}.}
\begin{equation}
n_{\log m} \: \propto
\begin{array}{ll} 
m^{-0.5}    &   \hbox{  for } 0.1<m<0.5 \\
m^{-\Gamma} &   \hbox{  for } 0.5<m<120 \\
\end{array}
\label{eqn:imf}
\end{equation}
where $n_{\log m} \, \dd \log m$ is the number of stars with logarithm
of the mass in the range $\log m$ to $\log m + \dd \log m$, $m$ is in
units of solar masses with 0.1\,\Msun\ and 120\,\Msun\ being the IMF
cutoffs. For the \cite{Salpeter55} IMF, $\Gamma=1.35$ in this
formalism except, traditionally, the slope continues down to
0.1\,\Msun.  A comparison between different IMFs is shown in
Figure~\ref{fig:imfs}.  Note that despite the average IMF of
\cite{Scalo98} using a break at 1 and 10\,\Msun, figure~5 of that
paper appears consistent with a single break at 0.5--0.8\,\Msun. Thus,
as with \cite{Kroupa01}, there is no strong evidence to rule out a
double-power law IMF with a break at 0.5\,\Msun\ being an adequate
approximation of a universal IMF. For the low-mass slope, we use the
average of the two \citeauthor{Kroupa01} equations.  We do not vary
this slope since our results are less sensitive to this than the
upper-mass slope.

\subsection{Chemical evolution}
\label{sec:model-chemical}

For chemical evolution, we incorporate the closed-box evolutionary
model within PEGASE which uses the models of \cite{WW95} to estimate
metallicity production via supernovae. The metallicity is controlled
by a parameter $r$ \citep{baldry02} that represents the total mass of
stars formed between $z=5$ and $z=0$ divided by the total amount of
gas initially available.  Higher values of $r$ produce higher
metallicity. The parameter can be greater than unity because of
recycling of material.  Since we are comparing different IMFs, we do
not quote $r$ values but rather the value for the metallicity
($\bar{Z}$) averaged on the luminosity at $z=0.1$ (the $\bar{Z}$-$r$
dependence varies with IMF).  Solar metallicity is considered to be
$Z=0.02$ in the PEGASE models.

\subsection{Dust attenuation}
\label{sec:model-dust}

We wish to estimate the effective dust attenuation for the cosmic
spectrum covering the UV to near-IR.  However, we can not consider a
standard slab or screen model as we would for an individual galaxy
because the contributions to the luminosity densities come from many
types of galaxies.  \cite{kochanek01} estimated 54\% and 46\%
contributions to the $K$-band from late- and early-type galaxies,
respectively.  The ratio is similar in the visible red bands from the
SDSS survey though it depends on the chosen dividing line in color or
morphology \citep{blanton03broadband} . From \cite{madgwick02}, 61\%
and 39\% are the fractional contributions to the \bj\ band from late
and early types, respectively, based on spectroscopic classification
(assuming Types~2--4 for late, Type~1 for early).  Late-type galaxies
(Sa--Sd and starbursts) contribute about 90--95\% of the light in the
0.2\,\micr\ UV \citep[estimated from table~7 of][]{sullivan00}.  This
is consistent with the results obtained by \cite{wolf03} for the
0.28\,\micr\ UV at $z=0.3$, with $\sim75/90$\% contribution from late
spectral types (3--4/2--4).  Using the inclination-averaged
attenuations of various galaxy types given by \citeauthor{Calzetti01}
(\citeyear{Calzetti01}, table~3) and averaging over suitable
distributions at each wavelength, we obtain cosmic spectrum
attenuations in magnitudes of 1.1--1.35 at 0.15\,\micr, 0.4--0.55 at
0.45\,\micr, 0.2--0.3 at 0.8\,\micr\ and 0.05--0.1 at 2.2\,\micr.

To incorporate dust attenuation, we use a power law. The attenuation
in magnitudes as a function of wavelength is approximated by
\begin{equation}
A_\lambda = A_v \left(\frac{\lambda}{\lambda_v}\right)^{-n}
\label{eqn:dust}
\end{equation}
where $A_v$ is the attenuation at the fiducial wavelength of
$\lambda_v$. We use 0.56\,\micr\ since it matches the effective wavelength
of our fiducial \rr\ band and it is close to the standard $V$
band (0.55\,\micr).

If we fit to the cosmic spectrum attenuations, estimated above, we
obtain $(A_v,n) \approx (0.35,1.0)$ for our fiducial dust model.  This
curve is within the ranges in attenuation estimated above at each
wavelength.  \cite{CF00} found that for an average star-burst galaxy,
$n$ was approximately 0.7 in a UV-to-visible attenuation law.  Their
model also included increased effective absorption in birth clouds
which had a finite lifetime.  Here, we assume that the the average of
all galaxies (the luminosity densities considered here) can be
approximated by a single effective optical depth. The average spectrum
of the Universe is not that of a star-burst galaxy and the broadband
colors are not strongly affected by emission lines.  The luminosity
density attenuations are naturally steeper than $n=0.7$ because of the
increasing contribution of less dusty ellipticals as the wavelength is
increased.  Rather than restricting our fitting to a single value of
$n$, we allow for a range from 0.8 to 1.2.

For $A_v$, we allow for a range from 0.2 to 0.55 magnitudes.  For
$n=1.0$, this is equivalent to a range in $A_{2000}$ from 0.55 to 1.55
magnitudes.  This encompasses the difference between the uncorrected
and dust-corrected luminosity densities of \citeauthor{sullivan00}\ 
which amounts to 1.3 magnitudes.  To avoid excessive attenuation at UV
and near-IR wavelengths, we also set $0.5\le A_{2000} \le1.6$ and 
$A_K\le0.15$ which reduces the $A_v$ range away from $n=1.0$.  In general,
we marginalize over attenuation, i.e., we chose the best fit
parameters within the defined ranges that minimize $\chi^2$ when
fitting the SFH and IMF.

\subsection{Summary of parameters}
\label{sec:summary-paras}

In this section, we summarize the important parameters and ranges
considered in our analysis and show the effect of varying some of the
parameters.
\begin{itemize}
\item Synthetic spectra. PEGASE.2 \citep{FR97,FR99}: Padova tracks
  \citep{bressan93}, spectral libraries of \cite{LCB97} and
  \cite{CM87}.  The \citeauthor{LCB97}\ library is principally derived
  from \cite{Kurucz92} model atmospheres. Nebular continuum and line
  emission is also included.
\item Cosmology. \cosmoparas\ = (0.7,0.3,0.7) (except luminosity
  densities are quoted for $h=1$, or, equivalently as $j+2.5\log h$).
  These values are fixed in our analysis. 
\item Cosmic SFH (Sec.~\ref{sec:model-sfh},
  Eqn.~\ref{eqn:cosmic-sfh}). $\beta$ between [0.5,4.0], $\alpha$
  between [$-$2,2], $\zform=5$. Low-redshift ($z<1$) star formation is
  parameterized by ${\rm SFR} \propto (1+z)^\beta$ and high-redshift
  by $(1+z)^\alpha$.
\item Universal IMF (Sec.~\ref{sec:model-imf}, Eqn.~\ref{eqn:imf}).
  Power-law slopes of $-0.5$ (for $m<0.5$\,\Msun) and $-\Gamma$
  between $[-0.8,-1.8]$ (for $m>0.5$\,\Msun) with $(m_{\rm min},m_{\rm
    max})$ = (0.1,120). The power-laws are with respect to $\log m$ so
  that the Salpeter IMF slope is $-1.35$.
\item Chemical evolution (Sec.~\ref{sec:model-chemical}). Closed-box
  approximation with $\bar{Z}$ between [0.008,0.05] (\Zsun=0.02) from
  $r$ between [0.1,1.4].  The final luminosity-weighted metallicity
  $\bar{Z}$ is principally a function of $r$ and the IMF.
\item Dust attenuation (Sec.~\ref{sec:model-dust},
  Eqn.~\ref{eqn:dust}). Power-law slope of $-n$ between $[-0.8,-1.2]$
  with $A_v$ between [0.2,0.55]. The ranges are also constrained by
  $0.5\le A_{2000} \le1.6$ and $A_K\le0.15$.
\end{itemize}

\begin{figure*}[ht] 
\epsscale{2.0}\plotone{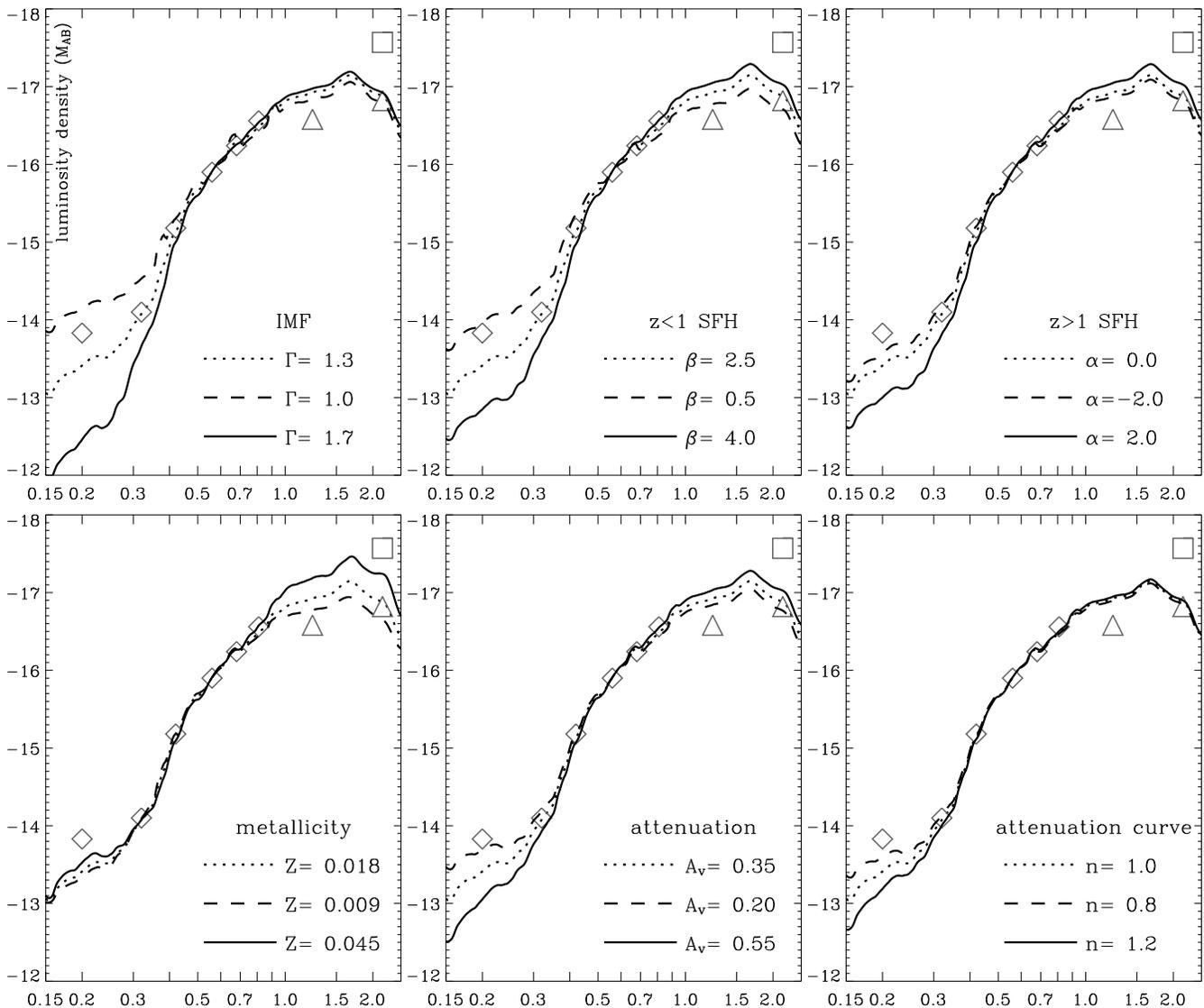}
\caption{The effect of varying selected parameters individually:{}
  absolute magnitude versus wavelength. The lines represent synthetic
  spectra smoothed to $\lambda/\Delta\lambda\sim10$. The dotted line
  is derived from the same set of parameters in every panel
  [$(\Gamma,\beta,\alpha,\bar{Z},A_v,n)$ = (1.3, 2.5, 0, 0.018, 0.35,
  1.0)]. All the spectra are normalized to $-15.90$ in the \rr\ band
  filter \citep{blanton03}. The symbols represent luminosity density
  measurements: the diamonds for \citeauthor{sullivan00}\ and
  \citeauthor{blanton03}; the triangles for \citeauthor{cole01}, and;
  the square for \citeauthor{huang03}}
\label{fig:example-variations}
\end{figure*}

Figure~\ref{fig:example-variations} shows the effects of varying one
parameter on the synthetic spectra (normalized at the \rr\ band) with
respect to a model of $(\Gamma,\beta,\alpha,\bar{Z},A_v,n)$ = (1.3,
2.5, 0, 0.018, 0.35, 1.0).  Over our chosen parameter ranges, varying
the IMF slope $\Gamma$ has the largest effect on the UV\,$-$\,\rr\ 
colors, with $\beta$ having the second largest effect.  The
metallicity has the largest effect on the near-IR\,$-$\,\rr\ colors.

\section{Results}
\label{sec:results}

\begin{figure*}[ht] 
\epsscale{1.0}\plotone{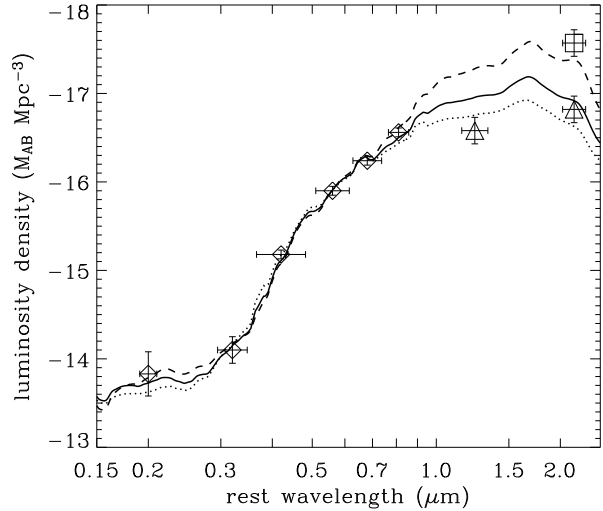}
\caption{Example fits to data{} 
  from SBC$_K$ (UV, $ugriz$ and $K$ triangle; solid line), SBH (UV,
  $ugriz$ and $K$ square; dashed line) and SBC$_J$ (UV, $ugri$ and
  $J$; dotted line).  The lines represent the fitted synthetic spectra
  smoothed to $\lambda/\Delta\lambda\sim10$.  The parameters for these
  fits were $(\Gamma,\beta,\alpha,\bar{Z},A_v,n)$ = (1.1, 4.0, 0,
  0.022, 0.20, 0.9) for SBC$_K$, (1.1, 3.5, 0, 0.048, 0.44, 0.8) for
  SBH and (1.1, 4.0, 0, 0.009, 0.22, 0.8) for SBC$_J$.  See text for
  details.}
\label{fig:example-fits}
\end{figure*}

First, we fit to the UV-to-optical luminosity density measurements of
\cite{sullivan00,blanton03} with different near-IR measurements,
considered separately: (a) using the \cite{cole01} $K$-band result
(compilation designated SBC$_K$); (b) using the \cite{huang03} result
(compilation SBH), and; (c) using the \citeauthor{cole01}\ $J$-band
result (compilation SBC$_J$) and not including the $z$-band
measurement which is included in the first two compilations.  Examples
of fitted spectra are shown in Figure~\ref{fig:example-fits}.  This
shows that the near-IR luminosity density measurements can be fitted,
primarily by varying metallicity, while the UV-to-optical spectrum
remains approximately the same.  Within the range of metallicity
considered here (0.4\,\Zsun--2.5\,\Zsun), the best fit is obtained to
the SBC$_K$ data.  Here, all the measurements are within about
$1\sigma$ of the synthetic magnitudes.  The new results of
\cite{blanton03} and this analysis resolves the major discrepancy
noted by \cite{Wright01} between the optical and near-IR luminosity
densities.  However, there is a significant discrepancy between the
Hawaii and the 2MASS $K$-band luminosity densities and a discrepancy
between the $J$-band result and the $z$-band/$K$-band results.

Neither of the infrared surveys is ideal for comparison with the SDSS
survey results. The 2MASS survey is too shallow (median $z\sim0.05$ at
the limit of the extended source catalog) for direct comparison with
the $z=0.1$ analysis and the Hawaii survey goes deeper but has a
significantly smaller area (8\,\sqdeg). Note that the
\citeauthor{sullivan00}\ result only uses about 200 galaxies but it is
consistent with the $^{0.1}u$ band result. Improved surveys are needed
to resolve the near-IR luminosity density discrepancies and to reduce
the uncertainties in UV luminosity densities.

\begin{figure*}[ht] 
\epsscale{1.0}
\plotone{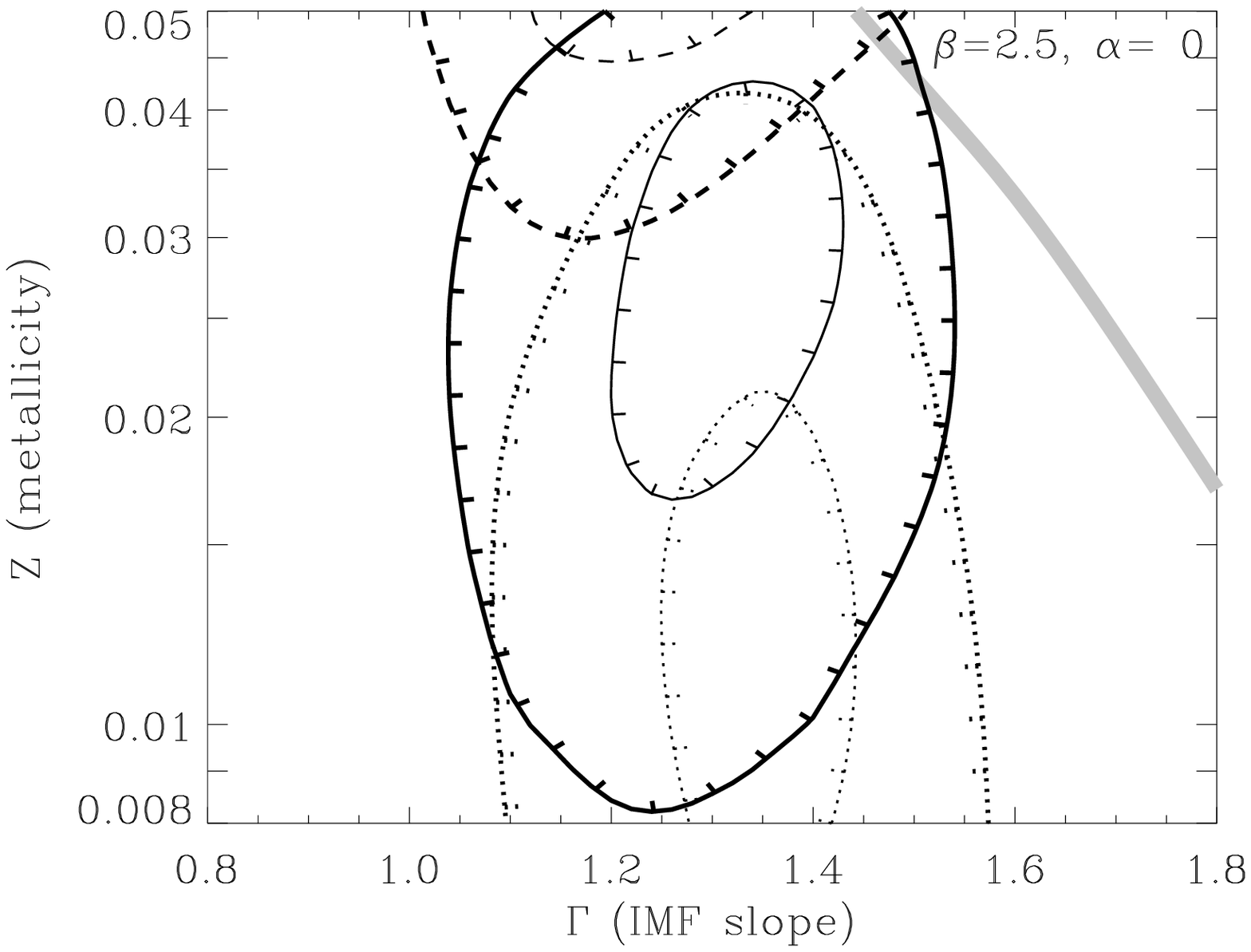}
\plotone{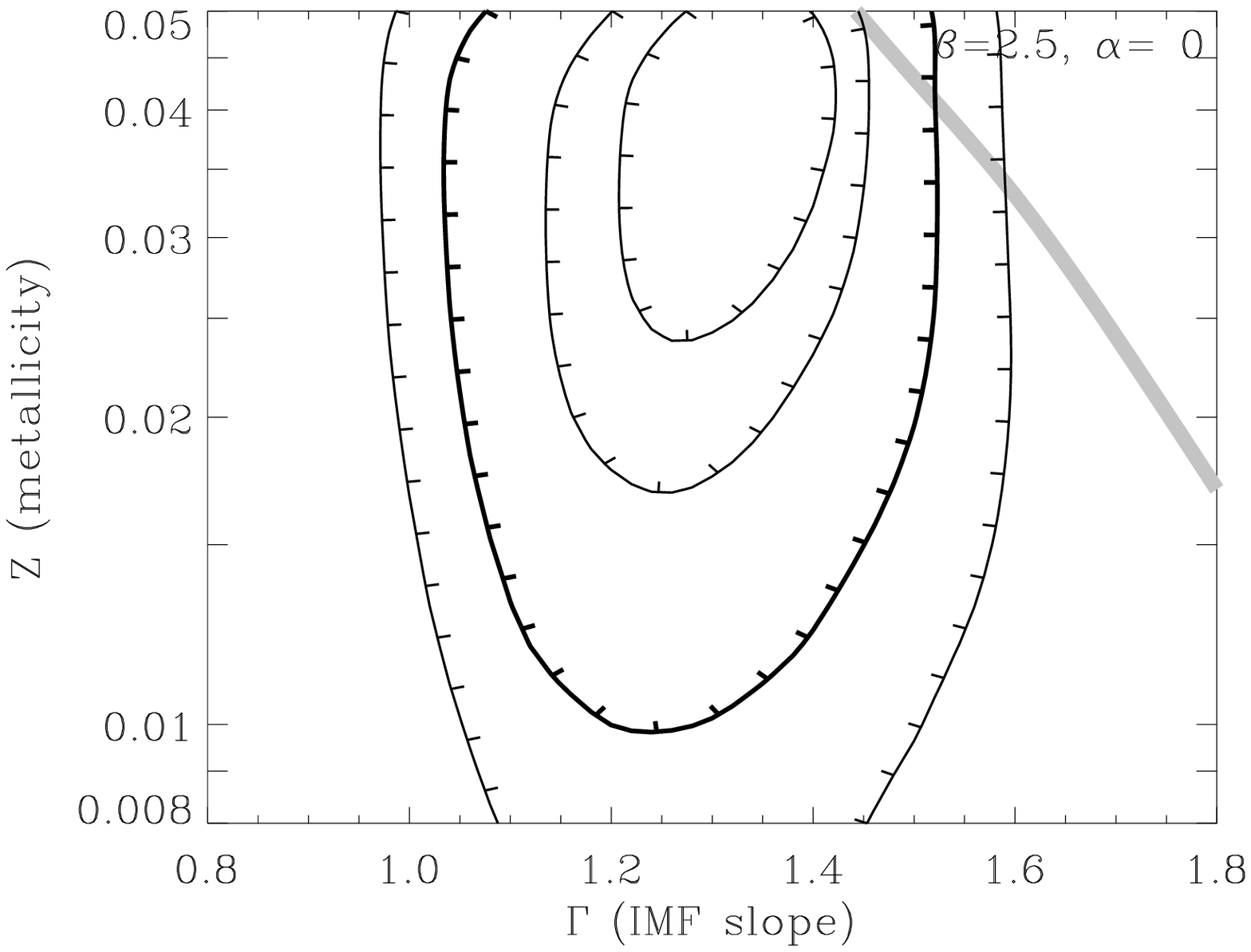}
\caption{Joint confidence levels over metallicity and IMF slope{} 
  assuming a cosmic SFH ($\beta=2.5$, $\alpha=0$). {\bf Left:} The
  contours represent confidence levels for the SBC$_K$ data (solid
  lines), SBH data (dashed lines) and SBC$_J$ data (dotted lines).
  The levels correspond to 68\% 1-para.\ and 95\% 2-para.\ confidence
  limits, marginalized over dust attenuation. The thick gray line
  represents the approximate upper limit on the luminosity-weighted
  $\bar{Z}$ due to chemical evolution (from the closed-box model of
  PEGASE). Any $\chi^2$ values above this line are linearly
  extrapolated.  {\bf Right:} The same except the confidence limits
  are for the SBav data which uses an average near-IR luminosity
  estimate (Eqn.~\ref{eqn:average-K}), and the levels represent the
  68\% 1-para.\ and 68\%, 95\%, 99.7\% 2-para.\ confidence limits.}
\label{fig:Z-imf+0}
\end{figure*}

We first look at constraints on the metallicity and IMF for a given
cosmic SFH (marginalizing over dust attenuation).  We compute $\chi^2$
and determine the confidence levels of $\Delta\chi^2$ = (1.0, 2.3,
6.2, 11.8) corresponding to (68.3\%) 1-parameter and (68.3\%, 95.4\%,
99.73\%) 2-parameter confidence limits. The resulting contours are
shown in Figure~\ref{fig:Z-imf+0} for a cosmic SFH with $\beta=2.5$
and $\alpha=0$ \citep{Hogg02,steidel99}.  As expected, the best fit
metallicity is dependent on the near-IR data: the SBC$_K$ data has a
best fit metallicity around 1--1.5\,\Zsun; SBH data, $>2$\,\Zsun, and;
SBC$_J$ data, $<1$\,\Zsun. The SBC$_K$ set of data is in agreement
with solar neighborhood measurements of the metallicity which give an
average close to solar \citep{Haywood01}.  The SBH data is in
disagreement with solar metallicity at the 99.7\% confidence level.

For the rest of the paper, we use an average estimate of the $K$-band
luminosity corresponding to 
\begin{equation}
j=-16.88\pm0.25
\label{eqn:average-K}
\end{equation}
from \cite{cole01} and \cite{kochanek01}. The uncertainty was
increased, and the $J$-band was not included, so that the constraints
on the IMF were not dependent on the discrepancies noted above. This
1$\sigma$ range in the $K$-band luminosity density is also similar to
the range determined by \cite{bell03}. The metallicity-IMF contours
for this data set (compilation designated SBav) are also shown in
Figure~\ref{fig:Z-imf+0}.  The best fit IMF slope, $\Gamma=1.3\pm0.1$,
shows the tightness of the constraint if the SFH is known.  The
best-fit metallicity is greater than solar but for other cosmic SFHs
(e.g.\ $\beta=2.5$ and $\alpha=2$), the best fit is around solar.  In
general, there is minimal degeneracy between $\Gamma$ and metallicity
for the data, i.e.\ the choice of metallicity does not significantly
affect the constraints on the IMF slope.

Note that there is an upper limit on the metallicity as a function of
$\Gamma$ due to the closed-box model (Fig.~\ref{fig:Z-imf+0}).
Insufficient metals can be produced in the available time to raise the
average metallicity above a certain limit and this limit decreases as
the fraction of high-mass stars in the IMF decreases. This chemical
evolution limit is derived from the PEGASE code using the upper
yields of \cite{WW95}.

\begin{figure*}[ht] 
\epsscale{1.0}
\plotone{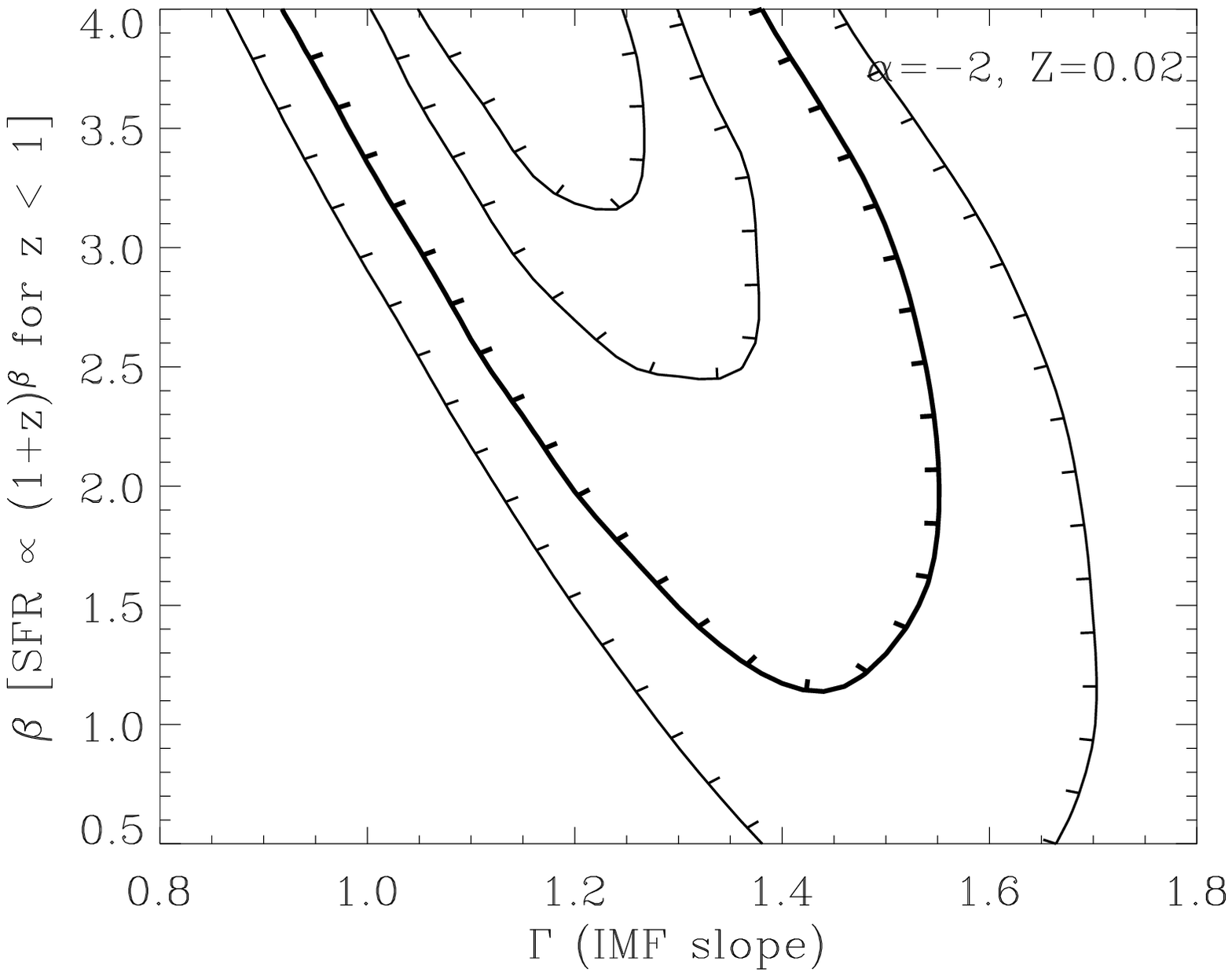}
\plotone{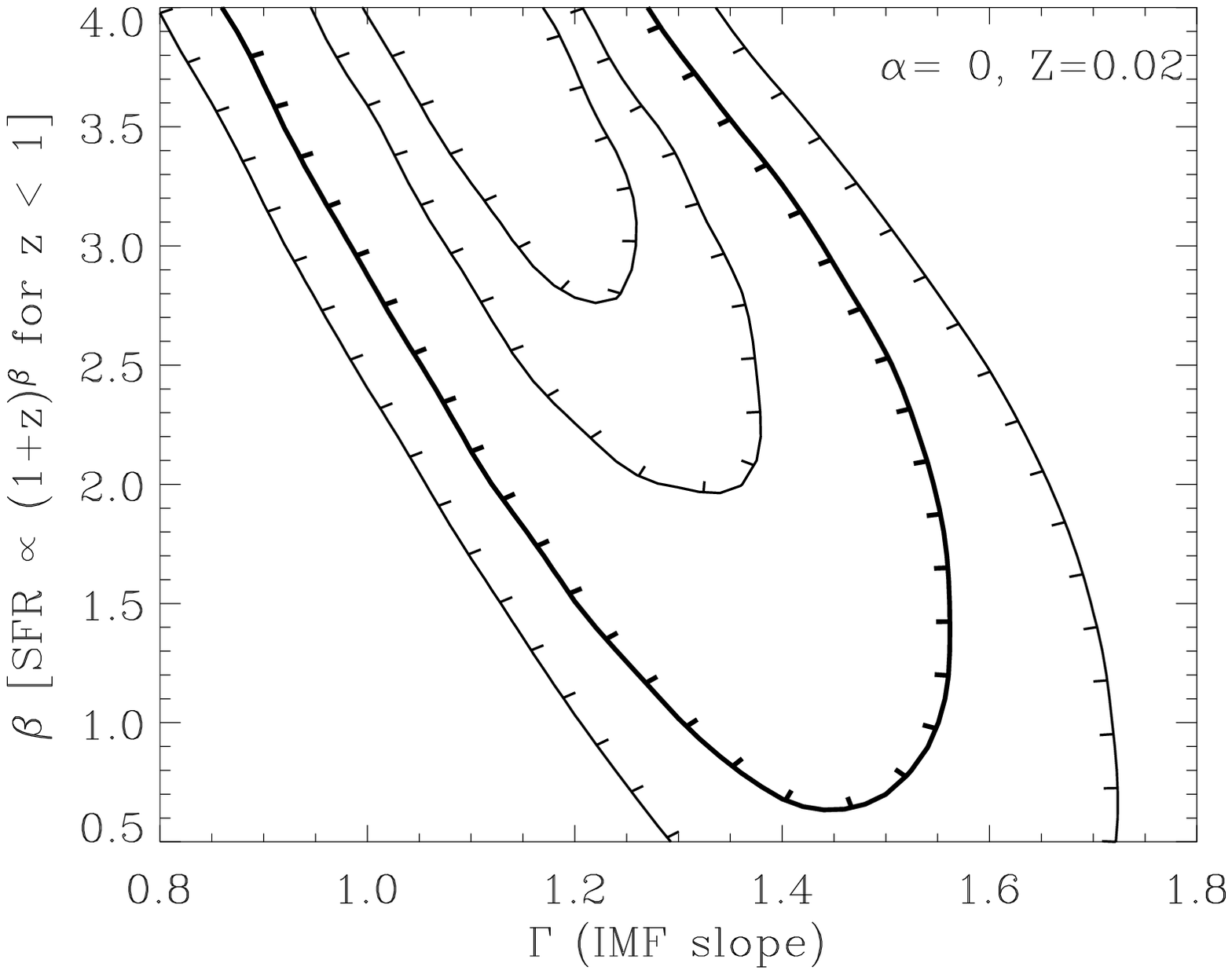}
\plotone{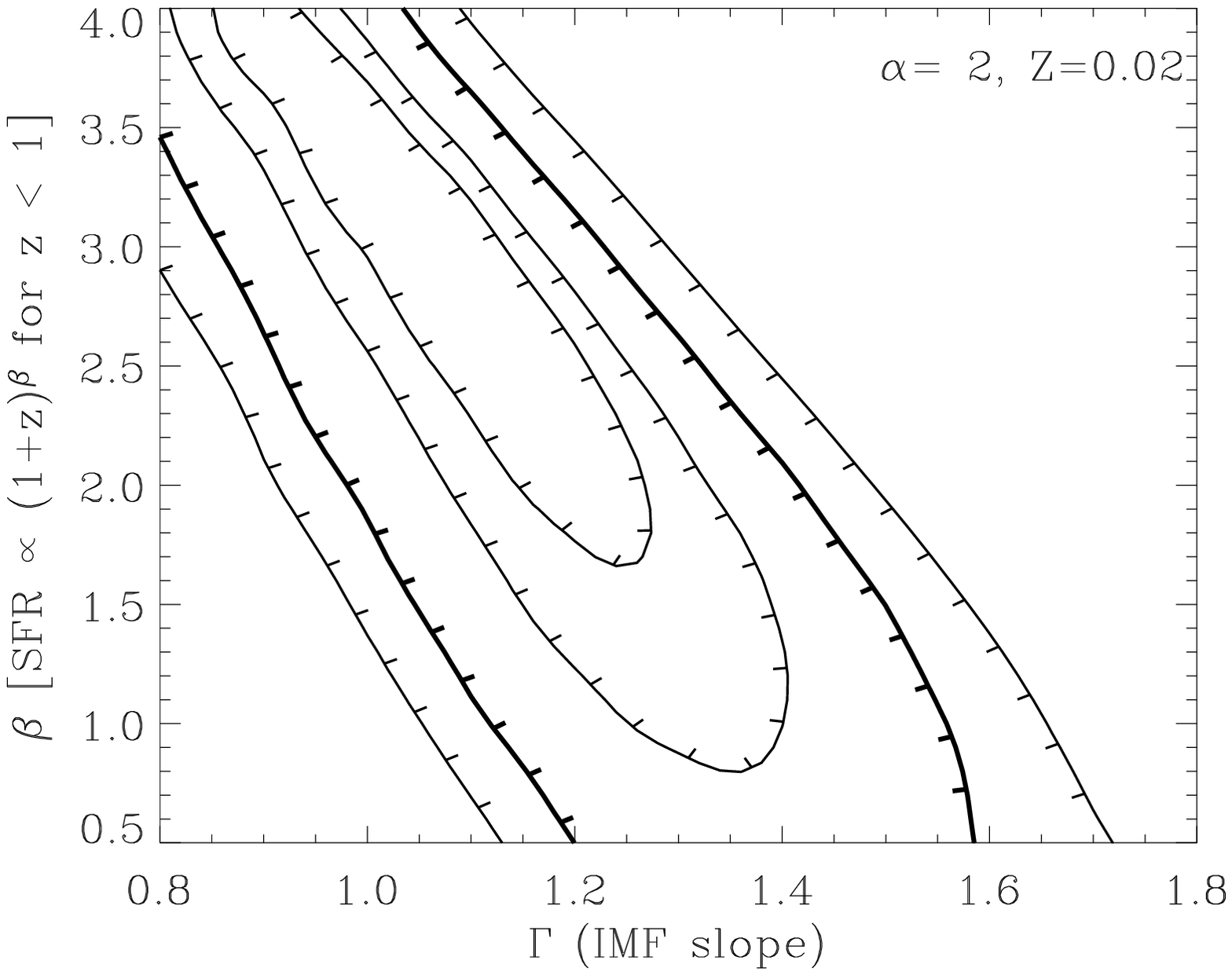}
\plotone{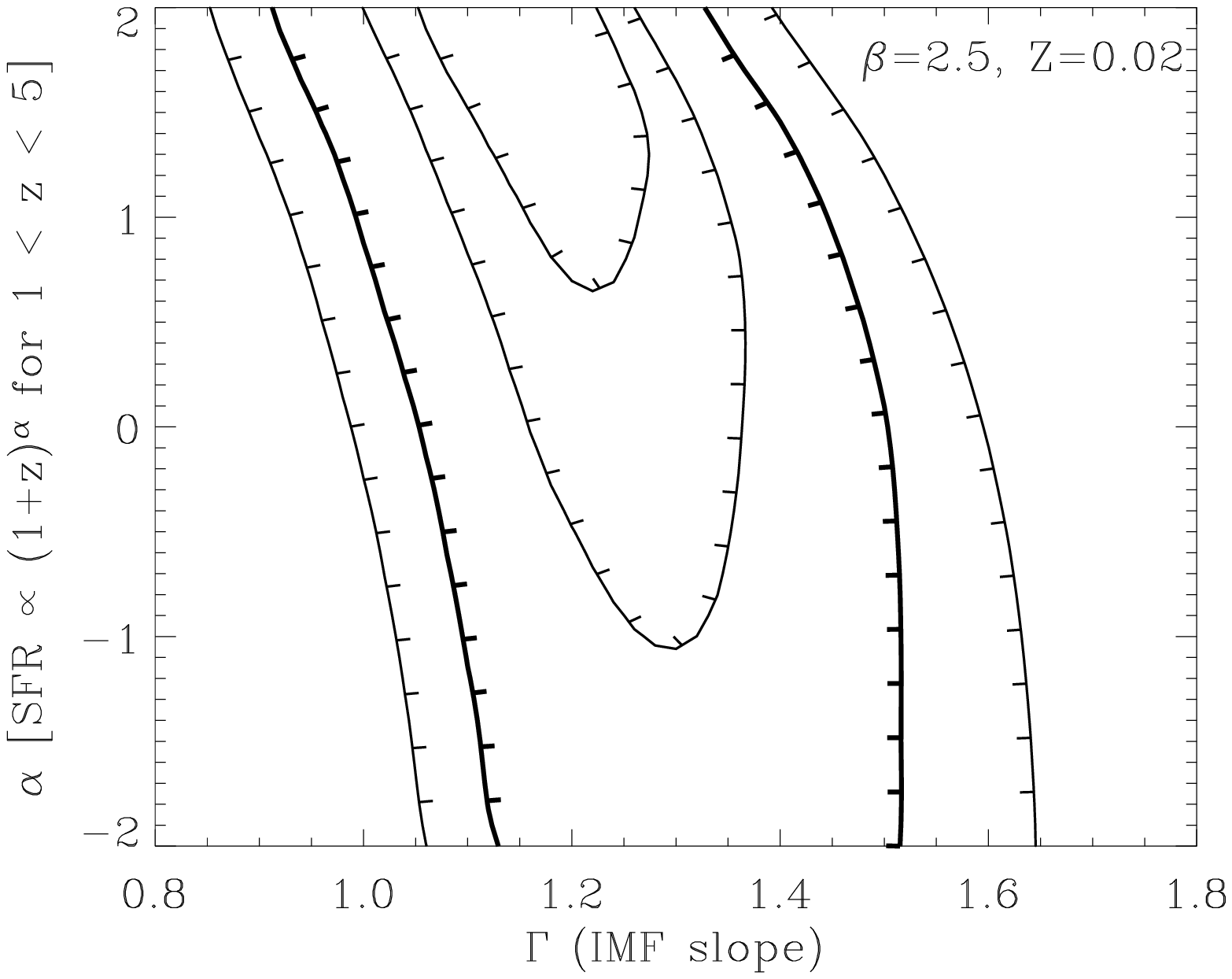}
\caption{Joint confidence levels over $\beta$/$\alpha$ and IMF slope{} 
  for three different values of $\alpha$ ($-$2,0,2) / one value of
  $\beta$ (2.5) and for a chemical evolution closed-box approximation
  with $\bar{Z}=0.02$.  The contours represent 68\% 1-para.\ and 68\%,
  95\%, 99.7\% 2-para.\ confidence levels for the SBav data,
  marginalized over dust attenuation.}
\label{fig:beta-imf}
\end{figure*}

The pattern shown in Figure~\ref{fig:Z-imf+0} is similar as $\beta$ is
varied except the best-fit $\Gamma$ shifts. To illustrate this, we now
fix $\bar{Z}$ equal to 0.02 for the SBav fit and show joint confidence
levels in $\beta$ versus $\Gamma$ in Figure~\ref{fig:beta-imf}. We
plot the result for three values of $\alpha$. All the plots show a
broad degeneracy with a slope of $-7\pm1$. In other words, the
best-fit $\Gamma$ decreases by about $-0.14$ for each $+1$ increase in
$\beta$. Note that the metallicity was chosen to be consistent with
the solar neighborhood and so that the closed-box model was valid out
to $\Gamma$ of 1.8 with only minimal extrapolation (see
Fig.~\ref{fig:Z-imf+0}). 

The choice of high redshift star formation (defined by $\alpha$) makes
little difference on the best-fit $\Gamma$ (Fig.~\ref{fig:beta-imf}).
Marginalizing over SFH ($\beta$ \& $\alpha$), we obtain a best fit for
$\Gamma$ in the range 0.85--1.3 (68\% conf.) and a strong upper limit
of $\Gamma<1.7$ (99.7\% conf.).  This is our principal result. In
other words, assuming a declining SFR from $z=1$ to the present day
($\beta\ge0.5$) then the present luminosity densities, in particular,
the UV to optical colors mean that the upper-mass IMF slope cannot be
steeper than 1.7. The caveats are: (i) PEGASE evolutionary tracks and
stellar spectra are sufficiently accurate; (ii) cosmic SFH can be
approximated with the double power-law and the look-back time to $z=1$
is close to 7.7\,Gyr; (iii) there is a close to universal IMF with a
near unbroken slope above 0.5\,\Msun; (iv) chemical evolution can be
approximated by a single metallicity for each epoch with the
metallicity increasing with time (closed-box approximation); (v) the
average effect of dust can be approximated by a power-law within the
ranges considered; (vi) the Copernican principle, that we are in no
special place in the Universe, applies; (vii) the luminosity density
uncertainties can be approximated as Gaussian.

We now turn to look at varying a couple of these assumptions, related
to metallicity and dust, in Sections~\ref{sec:metal-approx}
and~\ref{sec:dust-approx}. In Section~\ref{sec:further-densities}, we
calculate the stellar mass, SFR and bolometric densities from our
models and, in Section~\ref{sec:h-alpha}, we apply constraints using
measurements of H$\alpha$ luminosity density.

\subsection{Metallicity approximations}
\label{sec:metal-approx}

We have used the `closed-box' approximation for the evolutionary
synthesis in the above analysis.  This is valid if the dominant star
formation at each epoch is taking place in an average chemical
environment at that epoch. This scenario could result from complete
mixing between different environments.  Hot, young stars have a higher
metallicity on average than cool, old stars.  Another commonly used
assumption in evolutionary synthesis is the `constant-metallicity'
approximation. This scenario could result from no mixing between
environments. The metallicity evolves independently and rapidly in
each separate star-forming region with a characteristic or average
metallicity representing all epochs (separate regions) of star
formation.

\begin{figure*}[ht] 
\epsscale{1.0}\plotone{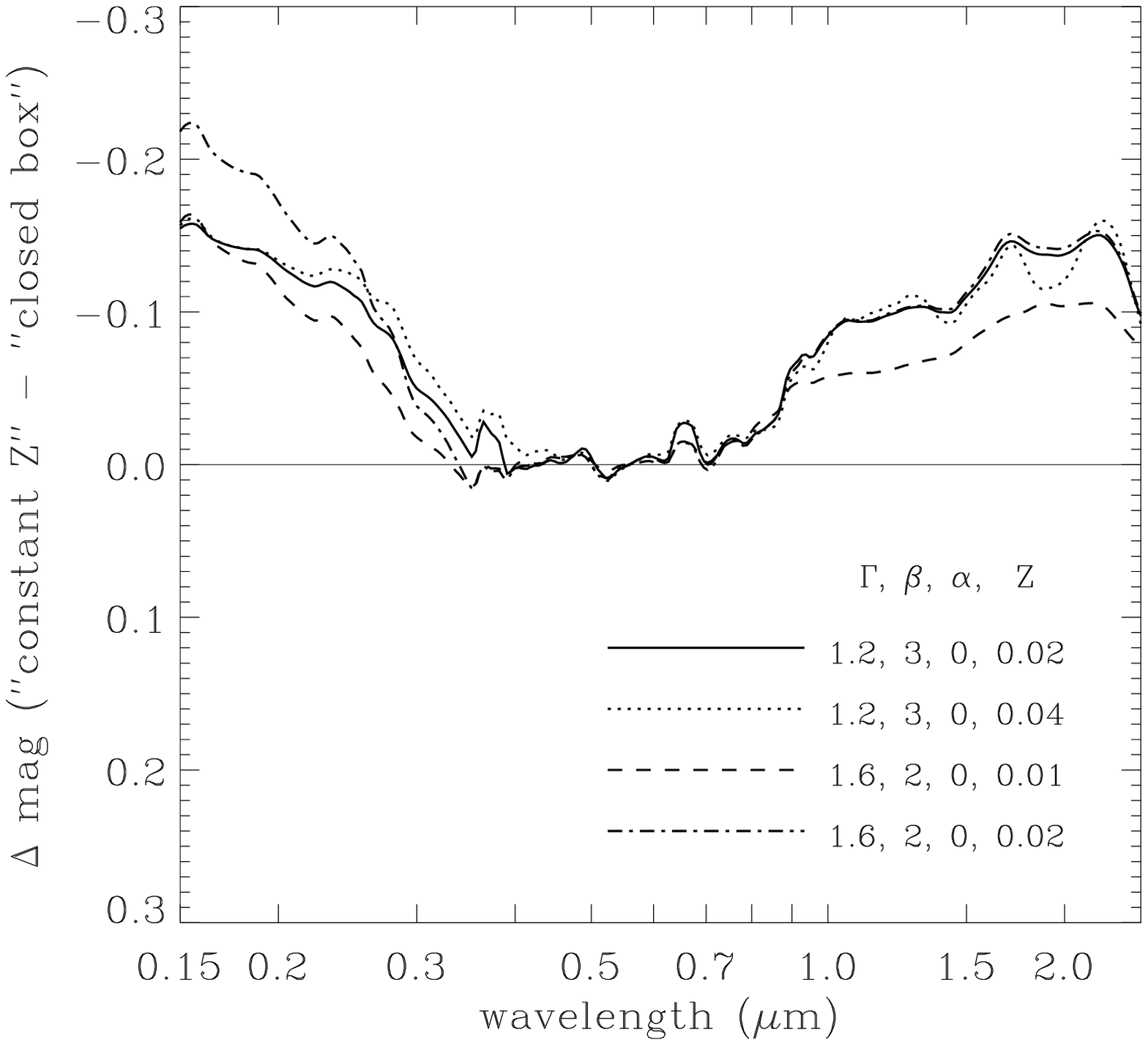}
\caption{Examples of the difference between constant-Z and 
  closed-box approximations{} on synthesized colors (wrt.\ the
  \rr\ band). Both the UV $-$ visible and near-IR $-$ visible
  fluxes are increased for the constant-Z approximation. The
  differences are plotted for matching luminosity-weighted $\bar{Z}$
  to constant $Z$.}
\label{fig:Z-approximations}
\end{figure*}

Since the Universe has neither complete or no mixing, something
between these two approximations might be expected.\footnote{We do not
  consider infall or outflow models which are often considered for
  individual galaxies.} Examples of differences between these are
shown in Figure~\ref{fig:Z-approximations}. Though the effect on the
broadband colors depends on the SFH and IMF, the general effect is to
increase the UV and near-IR fluxes with respect to the visible fluxes
for the constant-$Z$ approximation. This approximation leads to lower
metallicities for young/hot stars and higher metallicities for
old/cool stars compared to the closed-box evolution scenario (matching
$\bar{Z}$ to constant $Z$).

\begin{figure*}[ht] 
\epsscale{1.0}\plotone{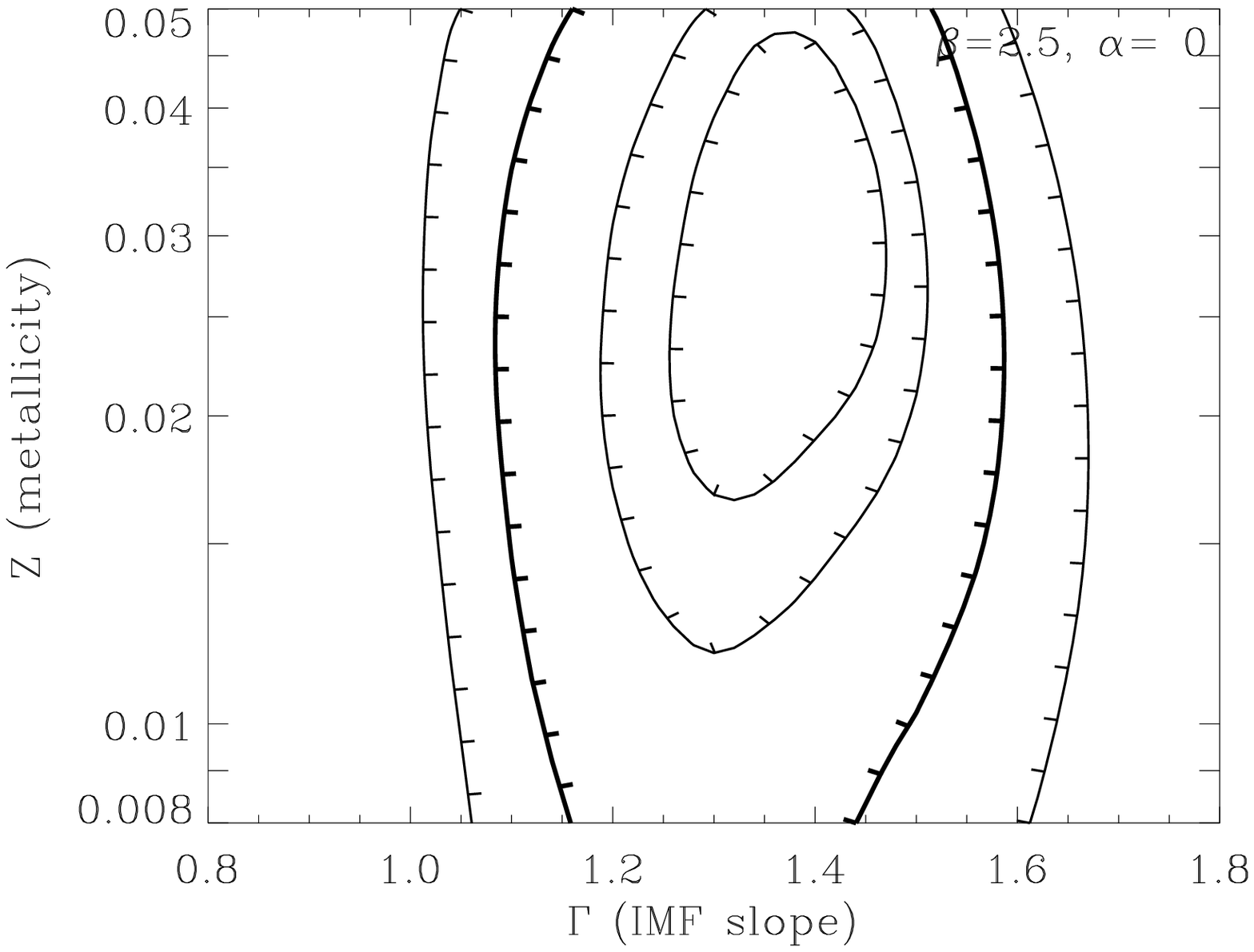}
\caption{Joint confidence levels over metallicity and IMF slope{} 
  assuming a cosmic SFH ($\beta=2.5$, $\alpha=0$) for the
  constant-metallicity approximation.  See Figure~\ref{fig:beta-imf}
  for contour meanings.}
\label{fig:const-Z-conf-A}
\end{figure*}

\begin{figure*}[ht] 
\epsscale{1.0}
\plotone{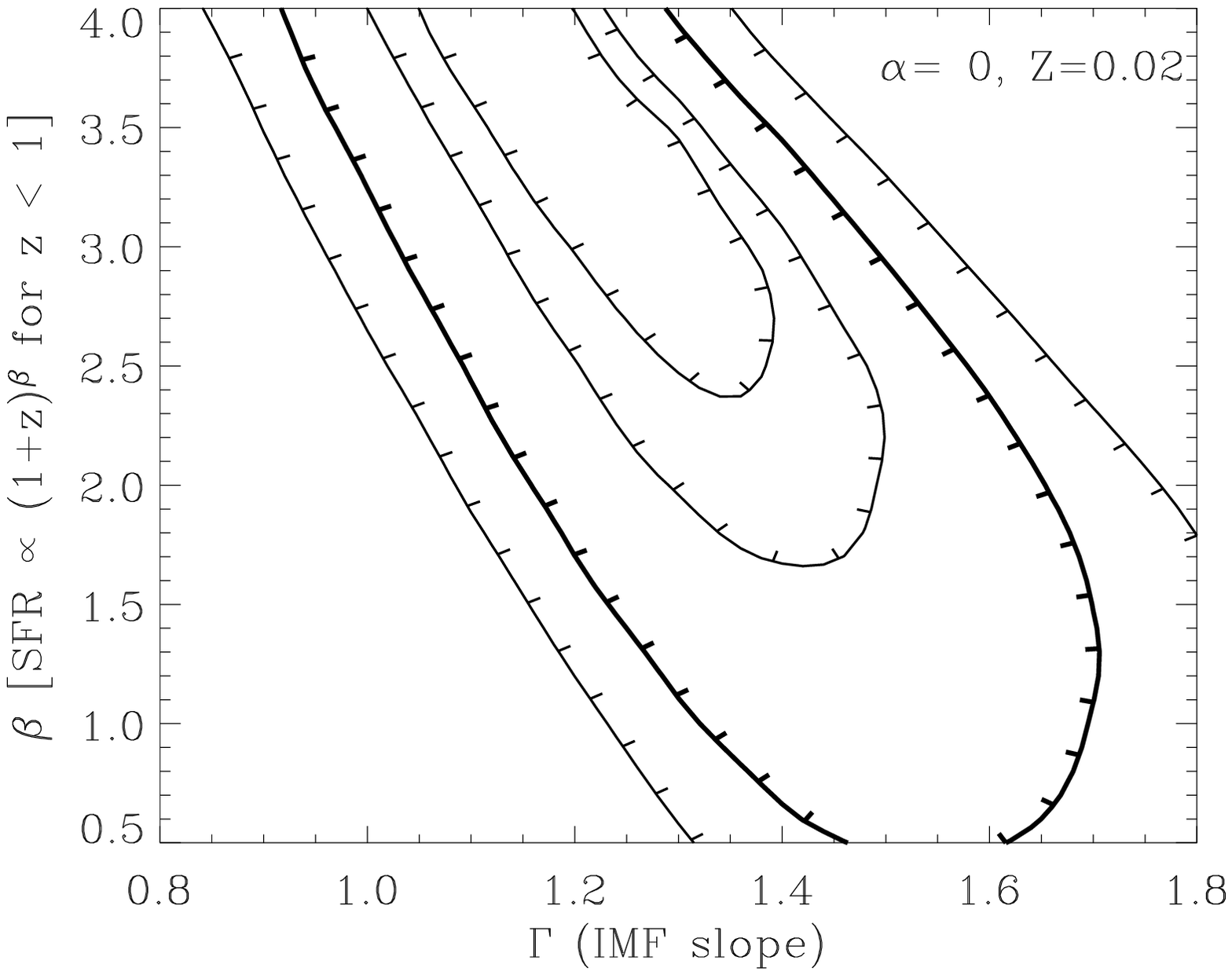}
\plotone{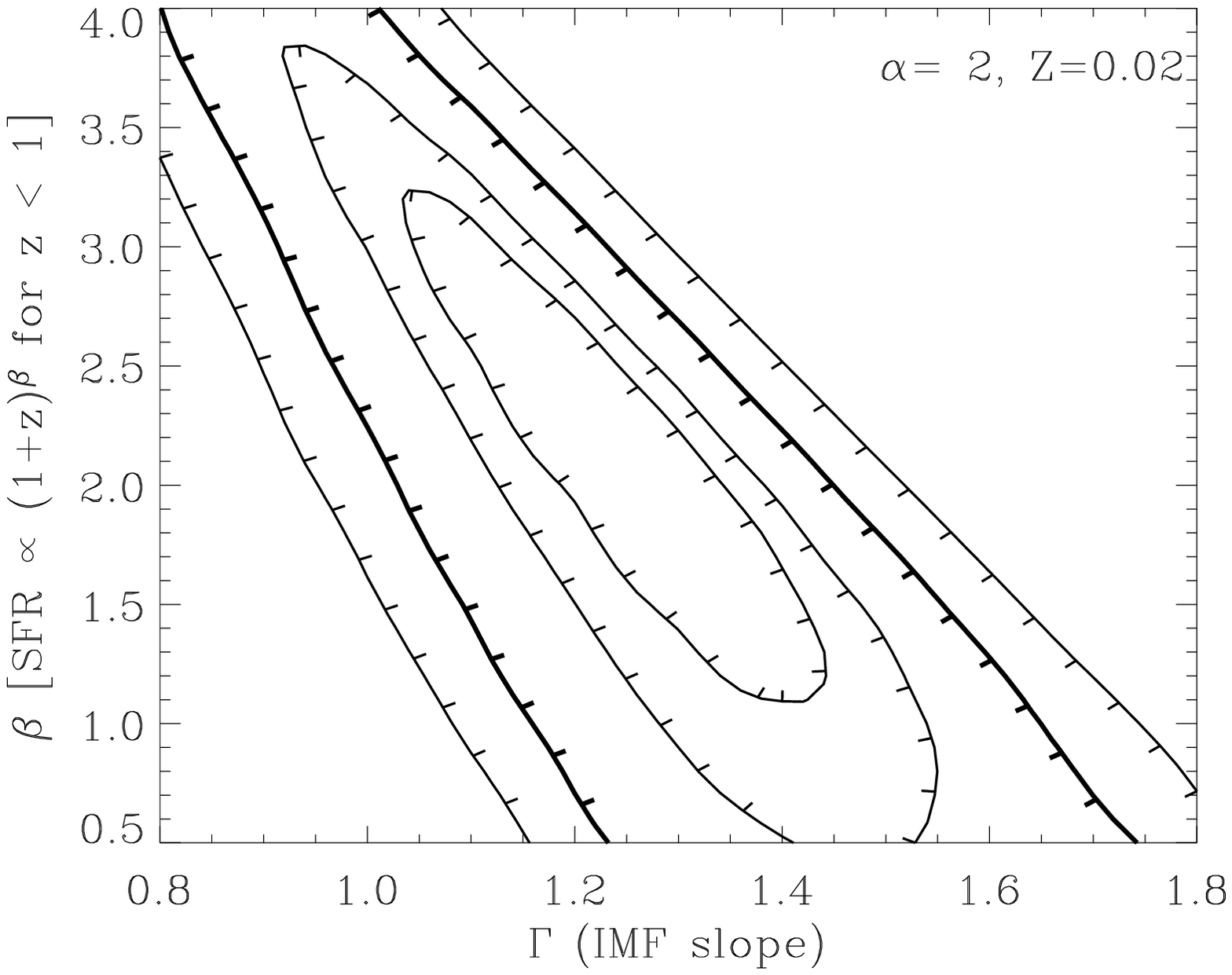}
\caption{Joint confidence levels over $\beta$ and IMF slope{} 
  for two different values of $\alpha$ (0,2) and for a
  constant-metallicity approximation with $Z=0.02$.  See
  Figure~\ref{fig:beta-imf} for contour meanings.}
\label{fig:const-Z-conf-B}
\end{figure*}

Some results for the constant-metallicity approximation are shown in
Figures~\ref{fig:const-Z-conf-A} and~\ref{fig:const-Z-conf-B} for the
SBav data. If we chose solar metallicity then the results are similar
to the closed-box approximation except $\Gamma$ is in the range
1.05--1.45 (68\% conf.) and $\Gamma<1.7$ at 95\% confidence. The
latter confidence level holds for $Z\la0.025$.

Table~\ref{tab:publ-imfs} shows a comparison between different
published IMFs with multiple power-law slopes.  We use the
constant-metallicity approximation for this comparison to avoid the
additional complication introduced by metal production in a closed-box
model.  With this comparison, the \cite*{KTG93,MS79,Scalo86} IMFs are
strongly rejected. This is consistent with their average slopes, over
1--10\,\Msun\ and 10--120\,\Msun, being $\Gamma\ge1.7$. Note that our
modeling cannot strongly distinguish between different slope changes
below 1\,\Msun.

\begin{table*} 
\caption{Comparison between published IMFs}
\label{tab:publ-imfs}
\begin{center}
\begin{tabular}{lllllll} \hline
reference& \multicolumn{4}{c}{\llap{$^a$}power-law slopes ($-\Gamma$) in four 
  mass ranges} & \llap{$^b$}confidence of & best-fit SFH $\beta$\\
&  0.1--0.5\,\Msun & 0.5--1\,\Msun & 1--10\,\Msun & 10--120\,\Msun
& rejection  & (for $\alpha=0$, 2) \\ \hline
this paper         & $-$0.5 &$-$1.2 &$-$1.2 &$-$1.2 & $<68$\%& 3.5, 2.5 \\
\cite{Kennicutt83} & $-$0.4 &$-$0.4 &$-$1.5 &$-$1.5 & 80\%   & 2.0, 1.0 \\
\cite{Kroupa01} A  & $-$0.3 &$-$1.3 &$-$1.3 &$-$1.3 & $<68$\%& 3.0, 2.0 \\
\cite{Kroupa01} B  & $-$0.8 &$-$1.7 &$-$1.3 &$-$1.3 & $<68$\%& 3.0, 2.0 \\
\cite{KTG93}       & $-$0.3 &$-$1.2 &$-$1.7 &$-$1.7 & 98\%   & 1.5, 0.5 \\
\cite{MS79}        & $-$0.4 &$-$0.4 &$-$1.5 &$-$2.3 & 98\%   & 1.5, 0.5 \\
Salpeter modified A& $-$0.5 &$-$1.35&$-$1.35&$-$1.35& $<68$\%& 3.0, 1.5 \\
Salpeter modified B& $-$0.5 &$-$0.5 &$-$1.35&$-$1.35& $<68$\%& 3.0, 2.0 \\
\llap{$^c$}\cite{Scalo86}
                   & $-$0.15&$-$1.1 &$-$2.05&$-$1.5 & 99.9\% & 0.5, 0.5 \\
\cite{Scalo98}     & $-$0.2 &$-$0.2 &$-$1.7 &$-$1.3 & 90\%   & 1.5, 0.5 \\
\hline
\end{tabular}
\end{center}
$^{a}$All the IMFs are assumed to be valid from 0.1\,\Msun\ to 120\,\Msun.
\newline
$^{b}$The IMFs were compared by marginalizing over 24 SFHs
($\beta=0.5$--4.0, step 0.5, for $\alpha=-2,0,2$) using the
constant-metallicity approximation with $Z=0.02$. The confidence of
rejection is with respect to the best-fit IMF in this table.\newline
$^{c}$The
power-law slopes shown for the \cite{Scalo86} IMF are an approximation
from a fit to the mass fractions (Fig.~\ref{fig:imfs}).
\end{table*}

\subsection{Dust attenuation approximations}
\label{sec:dust-approx}

We have used a power law to describe average dust attenuation based on
estimating the distributions of galaxy types at each wavelength
(Sec.~\ref{sec:model-dust}, $A_v=0.2$--0.55, $n=0.8$--1.2).
\cite{CF00} showed that star-forming galaxies were consistent with a
shallower or grayer slope of $n=0.7$.  If we take the approximate
distribution in the attenuation parameter measured by
\cite{charlot02}, $A_v=0.8\pm0.3$, and we also take $n=0.7\pm0.1$,
both with normal distributions (cutoff $>0$), then the average
effective parameters over many galaxies are equivalent to ($A_v$,$n$)
$\approx$ (0.75,0.65).  In other words, there is a slight reduction in
the effective attenuation and a slight flattening of the curve because
fluxes are averaged rather than magnitudes. The attenuation for these
average parameters at 0.2\,\micr\ is 1.45 magnitudes, not far from the
average attenuation estimated by \cite{sullivan00} of 1.3. Their
attenuation corrections were based on Balmer line measurements with
conversion to UV attenuation using the \cite{Calzetti97} law.

\begin{figure*}[ht] 
\epsscale{1.0}\plotone{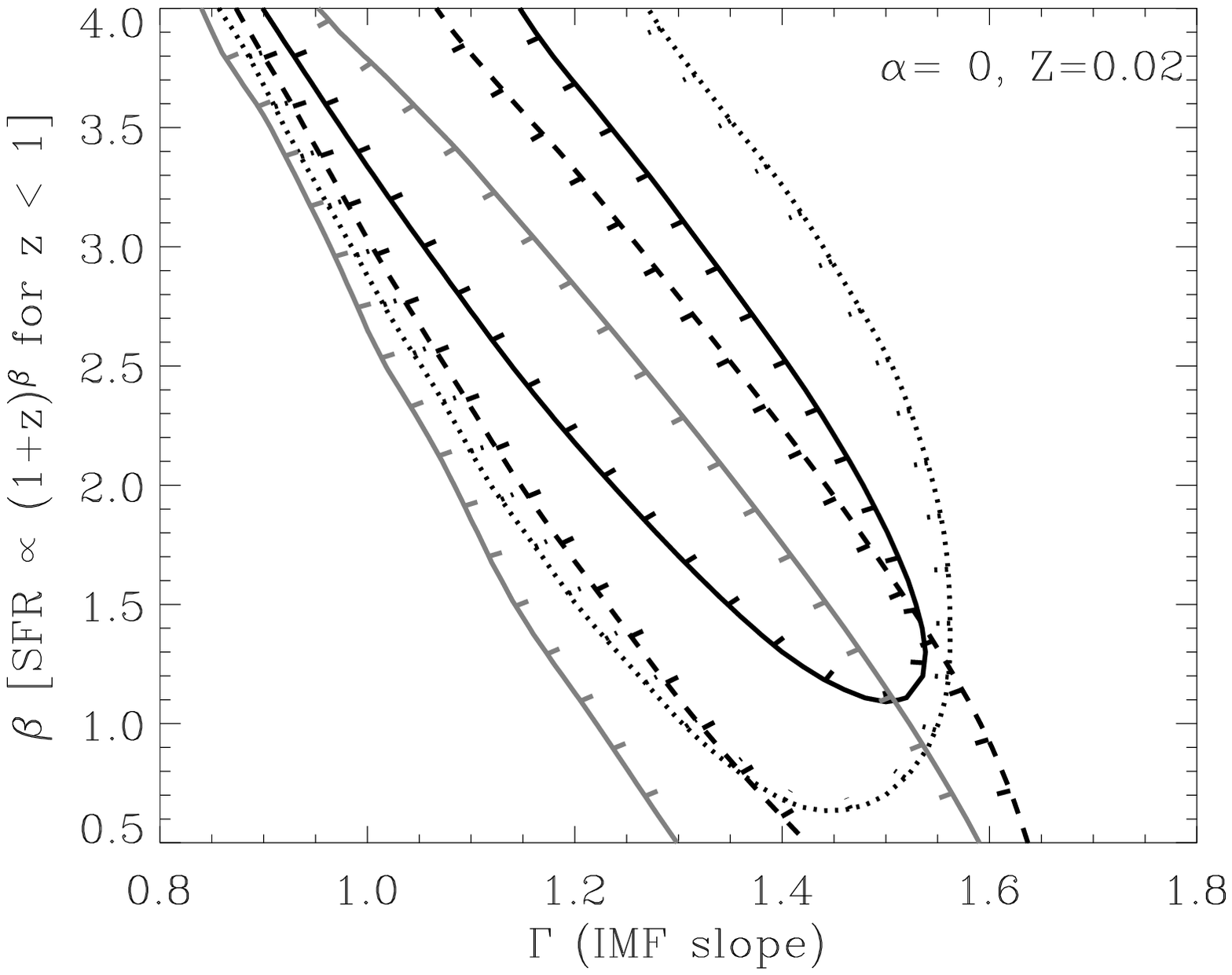}
\caption{Confidence levels in $\beta$ versus $\Gamma$ for various dust 
  approximations with the closed-box approximation. All the contours
  represent 95\% 2-para.\ confidence limits. The dotted line
  represents the result from marginalizing over attenuation
  $A_v=0.2$--0.55 and $n=0.8$--1.2 (Sec.~\ref{sec:model-dust}); the
  upper solid line from the fiducial $A_v=0.35$ and $n=1.0$ (effective
  average over many galaxies); the dashed line from fixed $A_v=0.75$
  and $n=0.65$ (star-forming galaxies), and; the lower solid line from
  fixed $A_v=0.5$ with an MW extinction law \citep{Pei92}.}
\label{fig:comp-dust}
\end{figure*}

In Figure~\ref{fig:comp-dust}, we compare this star-forming galaxy
attenuation law (0.75,0.65) with the average attenuation estimated in
Section~\ref{sec:model-dust} (0.35,1.0) and a steeper MW extinction
law \citep[$A_v=0.5$,][]{Pei92}. The best fit regions for these fixed
attenuation laws lie primarily within the best fit region based on
marginalizing over metallicity. Thus, our results are not strongly
dependent on the assumptions about dust and we have chosen a fairly
generous range of parameters for our principal fitting. Note that the
grayer law of (0.75,0.65) produces an unrealistically high attenuation
in the $K$-band of 0.3 magnitudes and the MW extinction law
($n\sim1.5$ visible-to-near-IR) is naturally too steep because it is a
foreground-screen law.  In Section~\ref{sec:further-densities}, we
analyse the consequences of our dust model for IR luminosity density
due to dust emission.

\subsection{Stellar mass, SFR and bolometric densities}
\label{sec:further-densities}

We can calculate some derived physical properties of the Universe (at
$z\approx0.1$).  The mass density in stars, \omstars, is in the range
1.1--2.0 $\times10^{-3}$ $h^{-1}$, and; the SFR density, \rhosfr, is
in the range 0.7--4.1 $\times10^{-2}$ \sfrunits. These ranges
represent 95\% confidence limits marginalized over IMF and cosmic SFH
but restricted to near-solar metallicity models ($\bar{Z}$ or
$Z=0.015$--0.025).  However, the results depend strongly on the
low-mass end of the stellar IMF which is not constrained by our
analysis.  As a test of the varying the low-mass end of the IMF, we
also computed the ranges using the best-fitting IMFs and cosmic SFHs
described in Table~\ref{tab:publ-imfs} (with $<95$\% conf.\ of
rejection).  From these, \omstars\ is in the range 0.8--2.5
$\times10^{-3}$ $h^{-1}$, and; \rhosfr\ is in the range 1.1--4.3
$\times10^{-2}$ \sfrunits.  The lower stellar mass densities are
derived from IMFs with $\Gamma_{m<1}\le0.5$ (Kennicutt, Salpeter
mod.~B, Scalo 1998) while the higher mass densities are derived from
the Kroupa (2001)~B IMF. The uncertainty in the current SFR density is
not increased compared to the measurement based on the
Equation~\ref{eqn:imf} parameterization. The lowest SFR densities
($<0.01$) only occur in our models with $\Gamma\la1.1$ and none of the
published IMFs in the table have $\Gamma_{m>1}<1.2$ which explains the
lack of low SFR densities based on those IMFs.

These results are in good agreement with the results, based on the
\citeauthor{Kennicutt83} IMF, of \citeauthor{cole01}\ and
\citeauthor{baldry02}, and with the results, based on a modified
Salpeter IMF, of \cite{bell03}.  They are generally not in agreement
with results based on the Salpeter IMF extending down to 0.1\,\Msun\ 
because such an IMF produces a high mass density from low mass stars
that have minimal impact on the luminosity densities.  This type of
universal IMF is ruled out by stellar counts in the MW
\citep[e.g.][]{Scalo86,Scalo98} and by analysis of the dynamics of
spiral galaxies \citep[e.g.][]{BdJ01}.

The total, bolometric, attenuated, stellar, luminosity density
(0.09--5\,\micr) is determined from the models to be in the range
1.2--1.7 $\times10^{35}$ \bolounits\ (95\% confidence).  We can also
estimate the total, bolometric, luminosity density
($\sim\,$5--1000\,\micr) due to dust absorbing and re-emitting stellar
light.  This depends critically on our dust model.  From the best-fit
models after marginalizing over dust-model parameters, the total is in
the range 0.3--1.5 $\times10^{35}$ \bolounits\ which corresponds to
20--50\% of the unattenuated stellar light being absorbed. If we
restrict our dust model to ($A_v$,$n$) = (0.35,1.0), the ranges are
0.55--0.95 $\times10^{35}$ \bolounits\ and 30--40\%.

From \cite{saunders90}, the local, far-IR, 42--122\,\micr, luminosity
density is in the range 0.17--0.26 $\times10^{35}$ \bolounits\ 
($\pm2\sigma$ range).  This is significantly lower than the energy
predicted by our dust model.  However, a correction to total dust
emission needs to be applied.  Using the infrared energy dust models
of \cite{DH02}, corrections from the 42--122 band emission to the
total dust emission range from 1.9 to 2.6. If we assume this
represents the systematic uncertainty in the luminosity density
correction, then the total bolometric dust emission is in the range
0.3--0.7 $\times10^{35}$ \bolounits\ (scaling from the
\citeauthor{saunders90}\ result).  Our fiducial dust model has a range
in total bolometric emission that overlaps with this estimate.

This upper limit of 0.7 $\times10^{35}$ \bolounits\ for the dust
emission favors models with lower attenuation and/or lower luminosity
densities around 0.2\,\micr. From our fitting, after marginalizing
over dust-model parameters, we obtain $A_{2000}\la1$. This is
marginally inconsistent with the \cite{sullivan00} estimate of the
effective attenuation, 1.3, which is why we have not used the
estimated total dust emission to constrain our dust model. This
discrepancy could be resolved if the UV luminosity density and/or
attenuation were overestimated,\footnote{However, recent analysis of
  the FOCA redshift survey with new redshifts and $k$-corrections
  gives a slightly more luminous 0.2\,\micr\ luminosity density by
  about 0.2 magnitudes (M.\ Sullivan 2003 private communication).}
and/or the total IR plus sub-mm flux was underestimated perhaps due to
a population of galaxies with colder dust than those detected by
60\,\micr\ surveys.  Note also that \cite{BB98} found an average
attenuation of 1.2 but this may not be inconsistent with our
attenuation limit since it represents a limit on the
luminosity-weighted average by flux.  Future analyses could use IR and
sub-mm luminosity density measurements to better constrain a dust
model.

\subsection{H$\alpha$ luminosity density}
\label{sec:h-alpha}

The H$\alpha$ nebular emission comes from reprocessed Lyman continuum
photons. Therefore it provides a measure of the UV flux blueward of
0.1\,\micr. Here, we consider measurements of the
attenuation-corrected H$\alpha$ luminosity density, relative to the
\rr\ band, in comparison with model predictions.  This emission-line
attenuation is significantly higher than for the stellar light at the
same wavelength \citep*{CKS94} but it can be estimated using the
Balmer decrement \citep[H$\alpha$/H$\beta \approx 2.85$ for case B
recombination,][]{HS87}.

\begin{table*} 
\caption{Attenuation-corrected H$\alpha$ luminosity densities}
\label{tab:h-alpha}
\begin{center}
\begin{tabular}{lcl} \hline
reference & $\log (L_{H\alpha}\,/\,h{\rm\,W\,Mpc}^{-3})$ & notes \\ \hline
\cite{gallego95}    & 32.39 & $z\approx0.03$, UCM objective-prism survey, luminosity function \\
\cite{TM98}         & 32.74 & $z\approx0.20$, CFRS, luminosity function \\
\cite{glazebrook03} & 32.77 & $z\approx0.03$, SDSS, cosmic spectrum \\ \hline
\end{tabular}
\end{center}
\end{table*}

Three attenuation-corrected H$\alpha$ luminosity density measurements
are summarized in Table~\ref{tab:h-alpha}. We do not quote error bars
as the uncertainties are dominated by systematics.  These include (i)
subtraction of the stellar contamination which, in particular, affects
the measurement of H$\beta$ and thus the estimate of the attenuation
(\citeauthor{glazebrook03}\ \citeyear{glazebrook03} included this
uncertainty which amounted to 0.1--0.15 in the $\log$ result), (ii)
uncertainties in the attenuation curve \citep[e.g.][]{CD86} which
affect the conversion from a reddening measurement to absolute
attenuation, and (iii) AGN contamination.  Both the \cite{gallego95}
and \cite{TM98} results are based on obtaining the emission-line
luminosity function from spectra. However, the selection criteria of
the surveys are different, an emission-line objective-prism survey and
$I$-band photometry, respectively.  \cite{glazebrook03} took a
different approach of summing SDSS spectra to form a cosmic optical
spectrum before calculating the H$\alpha$ and H$\beta$ luminosity
densities.  We take the mean of these three results for our fitting:
\begin{equation}
\log (L_{H\alpha}\,/\,h{\rm\,W\,Mpc}^{-3}) = 32.63\pm0.20 \: .
\label{eqn:h-alpha}
\end{equation}
This is appropriate since the \citeauthor{TM98} and
\citeauthor{glazebrook03}\ results may be too high due to AGN
contamination which are enhanced by aperture affects. The spectra
taken through a fiber are normalized to a broadband filter and since
AGN are centrally concentrated, any emission line luminosity due to
them will be over enhanced.  However, \citeauthor{glazebrook03}\ did
measure the weak O{\small I} 6300\,\AA\ line, suggesting that the AGN
contribution was only a few percent at most.  The
\citeauthor{gallego95}\ result is probably too low because of the
small survey (i.e.\ large scale structure) and/or because it only
includes EW$>$10\,\AA\ (H$\alpha$, emission positive) galaxies.

To compare this average measurement with model predictions, we use the
output from PEGASE for the H$\alpha$ flux without dust
attenuation.\footnote{Some fraction of Lyman continuum photons are
  assumed to be absorbed by dust rather than gas according to the
  prescriptions of \cite{Spitzer78}, normalized so that 30\% of the
  photons are absorbed by dust at solar metallicity \citep{FR99}.} As
with the earlier fitting, the synthetic spectra are normalized to the
\rr\ band so we are really comparing the ratio of H$\alpha$ and \rr\ 
band fluxes between the models and the data.  A correction is made for
the fact that the measured \rr\ band flux includes dust attenuation
because we are fitting to an attenuation-corrected H$\alpha$
measurement.  We use the fiducial attenuation of $A_v=0.35$.  In other
words, unlike the earlier fitting we now normalize the synthetic
spectra to an attenuation-corrected luminosity density in the \rr\ 
band.  

\begin{figure*}[ht] 
\epsscale{1.0}\plotone{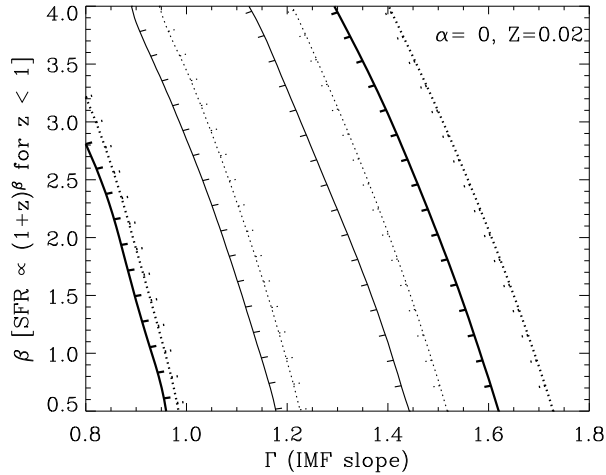}
\begin{center}
\caption{Confidence levels in $\beta$ versus $\Gamma$ from fitting
  solely to the H$\alpha$ luminosity density{} (Eqn.~\ref{eqn:h-alpha},
  relative to the \rr\ band). The solid lines represent the contours
  for the closed-box approximation while the dotted lines represent
  the contours for the constant-metallicity approximation.  The
  contours represent 68\% 1-para.\ and 95\% 2-para.\ confidence
  levels.}
\label{fig:cont-halpha}
\end{center}
\end{figure*}

Figure~\ref{fig:cont-halpha} shows the results from fitting to the
average attenuation-corrected H$\alpha$ luminosity density. The
degeneracy in $\beta$ versus $\Gamma$ is similar to that for the
broadband luminosity densities. In addition, the best-fit IMF slope,
$\Gamma$ in the range 0.9--1.5 (68\% conf.\ assuming $\alpha=0$ and
averaging over the metallicity approximations), is in good agreement
with the those results.

\section{Conclusions}
\label{sec:conclusions}

In this paper, we present the results of fitting spectral synthesis
models with varying IMFs to local luminosity densities.
\begin{itemize}
\item A good fit is obtained to the measurements of \cite{sullivan00},
  \cite{blanton03} and \cite{cole01} $K$-band (compilation SBC$_K$)
  with a best-fit metallicity of around solar.  If we fit to the
  measurements of the first two papers and \cite{huang03} (compilation
  SBH), the best fit metallicity is greater than twice solar and the
  results are inconsistent with solar metallicity at the 99.7\%
  confidence level.
\item The data can be well fit by a universal IMF and, therefore,
  there is no need to invoke IMF variations.  However, this provides
  only a weak constraint on the invariance of the IMF because of
  significant degeneracies associated with this type of modeling.
\item The best-fit universal IMF slope marginalized over a significant
  range of cosmic SFH ($0.5\le\beta\le4.0$, $-2\le\alpha\le2$) is
  $\Gamma=1.15\pm0.2$ based on an average between a closed-box
  approximation and a constant-metallicity approximation around solar
  (using compilation SBav).  Our results are in good agreement with
  the Salpeter IMF slope.
\item A strong upper limit of $\Gamma<1.7$ is obtained with 99.7\% or
  95\% confidence depending on the metallicity approximation. This
  rules out the \cite{Scalo86} IMF for a universal IMF since the mass
  fraction of stars above 10\,\Msun\ is similar to our
  parameterization with $\Gamma=1.8$ (Fig.~\ref{fig:imfs}, see also
  Table~\ref{tab:publ-imfs}). A similar conclusion was reached by
  \cite{MPD98} from fitting to luminosity densities over a range of
  redshifts and by \cite{KTC94} from fitting to H$\alpha$ EW-color
  relations for galaxies.  \citeauthor{MPD98}\ found a Salpeter IMF or
  an IMF with $\Gamma=1.7$ provided adequate fits. Here, we find that
  the latter slope does not provide a good fit to luminosity
  densities.  This is principally because we have used more accurate
  local luminosity density measurements.  If we increase our $1\sigma$
  uncertainties by 0.05 at all wavelengths then we obtain $\Gamma<1.7$
  with 80\% confidence.
\item The stellar mass density of the universe is in the range
  1.1--2.0 $\times10^{-3}$ $h^{-1}$ based on marginalizing over cosmic
  SFH and our parameterization of the IMF.  The current SFR density is
  in the range 0.7--4.1 $\times10^{-2}$ \sfrunits. The total
  bolometric stellar emission (0.09--5\,\micr) is known more
  accurately, naturally because we observe light and not mass, and is
  in the range 1.2--1.7 $\times10^{35}$ \bolounits\ derived from the
  fitted PEGASE model spectra (the mass-to-light ratio is
  $\sim0.9$--1.4 \Msun/\Lsun). We find that our dust model can
  reproduce the estimated total dust emission (0.3--0.7
  $\times10^{35}$ \bolounits, $\sim\,$5--1000\,\micr) scaled from the
  far-IR luminosity density \citep{saunders90} if we have
  $A_{2000}\la1$ ($\la\,$60\%). Note that this represents a limit on
  the cosmic spectrum attenuation, i.e., a luminosity-weighted average
  by flux (not magnitudes).
\item Fitting to the local H$\alpha$ luminosity density provides a
  similar result for the IMF slope ($\Gamma=1.2\pm0.3$). This
  provides some evidence that our upper mass cutoff of 120\,\Msun\ is
  a reasonable approximation because the sensitivity of the H$\alpha$
  flux to massive stars is different to that of the mid-UV to optical
  fluxes.  More accurate measurements could test this upper mass
  limit.  
\item The quantitative results on $\Gamma$ rely on the accuracy of the
  population synthesis model (PEGASE). An alternative, qualitative
  result would be that there is consistency between the theory of
  evolutionary population synthesis (evolutionary tracks and stellar
  spectra) and the measurements of luminosity densities, cosmic SFH
  and an average MW IMF derived from stellar counts \citep[e.g.\ 
  $\Gamma\approx1.3$ from][]{Kroupa01}.
\end{itemize}

Greater constraints can be placed on a universal IMF both by improved
accuracy in local luminosity density measurements (in particular,
$z=0.1$ to match the multi-wavelength SDSS MGS) and by improved
accuracy of direct measures of cosmic SFH with redshift (in
particular, $z=0$ to $z\sim1$--2).  For the direct tracing of cosmic
SFH, it is important that the UV is measured at the same rest-frame
wavelength in order to avoid IMF dependency. In other words, we need
an IMF-independent cosmic SFH in order for the local luminosity
densities to accurately constrain a universal IMF or the
IMF-dependency should be quantified.

\acknowledgements{We thank the referee Eric Bell for constructive
  suggestions, and also Timothy Heckman, Benne Holwerda, Rosemary Wyse
  and Andrew Blain for helpful comments. We acknowledge generous
  funding from the David and Lucille Packard foundation.}

\end{document}